\documentclass[review,12pt]{elsarticle}

\biboptions{numbers,sort&compress}

\usepackage{amssymb}
\usepackage{amsmath}
\usepackage[hidelinks]{hyperref}
\usepackage{url}
\urldef\urltdr\url{https://genxproject.github.io/GenX.jl/dev/Model_Reference/TDR/##Time-Domain-Reduction-(TDR)}

\journal{Environmental Research: Energy}

\begin{document}

\begin{frontmatter}



\title{Establishing best practices for modeling long duration energy storage in deeply decarbonized energy systems}


\author[mae,andlinger]{Gabriel Mantegna}
\ead{gabe.mantegna@princeton.edu}

\author[mae,andlinger]{Wilson Ricks}

\author[ece,andlinger]{Aneesha Manocha}

\author[suny]{Neha Patankar}

\author[nyu]{Dharik Mallapragada}

\author[mae,andlinger]{Jesse Jenkins}

\affiliation[mae]{organization={Department of Mechanical and Aerospace Engineering, Princeton University},
            city={Princeton},
            state={NJ},
            country={USA}}
            
\affiliation[andlinger]{organization={Andlinger Center for Energy and the Environment, Princeton University},
            city={Princeton},
            state={NJ},
            country={USA}}

\affiliation[ece]{organization={Department of Electrical and Computer Engineering, Princeton University},
            city={Princeton},
            state={NJ},
            country={USA}}

\affiliation[suny]{organization={Department of Systems Science and Industrial Engineering, Binghamton University},city={Binghamton},state={NY},country={USA}}

\affiliation[nyu]{organization={Department of Chemical and Biomolecular Engineering, New York University},city={Brooklyn},state={NY},country={USA}}

\begin{abstract}
Long duration energy storage (LDES) may become a critical technology for the decarbonization of the power sector, as current commercially available Li-ion battery storage technologies cannot cost-effectively shift energy to address multi-day or seasonal variability in demand and renewable energy availability. LDES is difficult to model in existing energy system planning models (such as electricity system capacity expansion models), as it is much more dependent on an accurate representation of chronology than other resources. Techniques exist for modeling LDES in these planning models; however, it is not known how spatial and temporal resolution affect the performance of these techniques, creating a research gap. In this study we examine what spatial and temporal resolution is necessarily to accurately capture the full value of LDES, in the context of a continent-scale capacity expansion model. We use the results to draw conclusions and present best practices for modelers seeking to accurately model LDES in a macro-energy systems planning context. Our key findings are: 1) modeling LDES with linked representative periods is crucial to capturing its full value, 2) LDES value is highly sensitive to the cost and availability of other resources, and 3) temporal resolution is more important than spatial resolution for capturing the full value of LDES, although how much temporal resolution is needed will depend on the specific model context.

\end{abstract}

%

\begin{keyword}
energy storage \sep long duration energy storage \sep capacity expansion \sep decarbonization \sep macro-energy systems


\end{keyword}

\end{frontmatter}


\section{Introduction}

Long-duration energy storage (LDES) may become a critical technology for enabling the deep decarbonization of the electric grid at reasonable cost. Cost-effective pathways to deeply decarbonize the electric sector are likely to involve very high levels of variable, weather-dependent renewable energy sources (e.g., wind and solar PV), which will increase the value of storage devices capable of economically shifting energy across weeks and/or seasons to align renewables output with demand. Commercially available Lithium-ion (Li-ion) storage devices are not techno-economically suited for this role \cite{mallapragada_long-run_2020}. The energy capacity cost necessary for a storage resource to be economic for multi-day storage is on the order of \$20 kWh$^{-1}$ and seasonal shifting requires on the order of \$1 kWh$^{-1}$, one-to-two orders of magnitude cheaper than the current installed costs of Li-ion battery systems ($\approx$\$300-400 kWh$^{-1}$) \cite{sepulveda_design_2021, cole_cost_2023}. There are several technologies in development that could serve this role with very low costs per kWh of energy stored, such as hydrogen with underground cavern storage, high-temperature thermal storage, or iron-air battery technology.

While LDES technologies may play a key role in decarbonizing the power sector, most energy system planning tools in use today are not capable of accurately capturing the role of LDES or the full value of these resources \cite{levin_energy_2023}. Most macro-energy system planning tools, including electricity system capacity expansion models (CEMs), use a reduced temporal resolution to represent operations, either with several unlinked representative hours, or with several unlinked representative operational periods, such as representative days or weeks. The latter is widely acknowledged to be a best practice for representing the need for intra-day energy storage, as it captures typical diurnal patterns of supply and demand \cite{bistline_importance_2021}; however, modeling unlinked representative periods cannot represent the full value of LDES specifically because it does not allow the shifting of energy \textit{between} representative periods, meaning it cannot capture multi-day or seasonal shifting \cite{levin_energy_2023}.

Kotzur et al. (2018) \cite{kotzur_time_2018} developed an algorithm that allows multi-day and seasonal energy storage behavior of LDES to be captured while still modeling system operations using representative periods (e.g. days, weeks). This general approach, which we refer to as the `period-linking method,' is described in further detail in Section \ref{sec:methods}. In brief, the period-linking method uses representative periods for operational modeling, but tracks the state of charge of LDES across an approximated version of a full year. This approach can also be thought of as decomposing the state of charge of LDES into intra-period and inter-period components, and modeling each separately. The period-linking method adds computational complexity by way of introducing new constraints that link or couple the representative periods, but the computational burden is minimal compared to that of modeling the operations of the entire system over the course of a year.

This period-linking algorithm represents a current best practice for LDES modeling when using representative periods. Similar methods have been implemented in macro-energy systems planning models such as GenX \cite{jenkins_enhanced_2017}, which is the model used in the current study, and RIO \cite{haley_supporting_2022}, and employed in several recent studies \cite{williams_carbon-neutral_2021, ricks_minimizing_2023, patankar_land_2023, baik_what_2021}. However, there remains a research gap regarding how to apply this period-linking method to accurately capture the full value of LDES. In particular, previous literature has not yet established how many representative periods are necessary, and how long these periods should be, in order to capture the full value of LDES. Developing and applying capacity expansion models involves many abstractions in order to reduce computational complexity, including temporal domain reduction, spatial domain reduction, generator clustering, and operational simplifications, and it is not well known which of these choices result in abstraction errors that significantly bias estimates of the value of LDES. This is also important because an inaccurate representation of LDES can lead to a model showing unrealistically high costs of emissions reductions, particularly for very low carbon energy systems, and such errors can distort capacity results for other resources, leading to an inaccurate picture of optimal resource portfolios. Capacity expansion models are widely used to inform investment decisions in the electric utility industry and to guide policymaking for the electric sector, so minimizing biases in these models is crucial.

To our knowledge, no prior work has examined the impact of model abstraction decisions on the value of LDES either: a) comprehensively (i.e., by changing multiple parameters at once so interactive effects can be explored), or b) in tandem with the period-linking algorithm, which represents a current best practice. A large body of literature has found that energy storage value in general is sensitive to model temporal resolution \cite{bistline_importance_2021, sioshansi_energy-storage_2022, liu_power_2022, abdulla_importance_2017, diaz_importance_2019, hoffmann_review_2020}, and several other studies have also found that capacity expansion models more generally are sensitive to spatial resolution \cite{martinez-gordon_review_2021, krishnan_evaluating_2016}, but none of these focus on LDES. Techniques such as representative days that are generally seen as a best practice for capturing short-duration intra-day storage capabilities such as Li-ion batteries are insufficient to accurately capture the value of LDES with capability to shift energy between representative periods, and none of these prior works examine the interactive effects of changing spatial and temporal resolution simultaneously. Two more recent studies examined the impact of temporal resolution on LDES value \cite{sanchez-perez_effect_2022, sanchez-perez_effect_2022-1}, but neither of these use the period-linking algorithm developed by Kotzur et al. \cite{kotzur_time_2018}, so their findings do not transfer over well to a situation in which this algorithm is used. The current study fills this gap in the literature by examining the impact of model abstraction choices on LDES value comprehensively, including spatial and temporal resolution, and by examining the question of temporal abstraction in a context where Kotzur et al.'s \cite{kotzur_time_2018} period-linking methodology is applied. Employing this period-linking methodology is important because this methodology allows modelers to simplify the time domain while still accurately modeling LDES, allowing them to increase model complexity in other areas, such as spatial resolution or operational details such as unit commitment.

The central research question of our study is: what model abstraction choices are necessary to accurately capture the value of LDES in a capacity expansion model? To answer this, we develop an electricity system capacity expansion model for the entire continental United States, and we vary model spatial and temporal abstraction to determine what combinations are sufficient to capture the full value of LDES. We model a ``generic'' 200-hr LDES resource with 42\% round-trip efficiency and use model results to estimate the marginal value of this technology at a specified deployment level (rather than choosing a specific technology with an assumed cost and estimate equilibrium capacity deployment). In addition to this core experiment of examining the impact of model abstraction on LDES value, we also test the robustness of our results to key dimensions of uncertainty such as cost and availability of competing technologies, stringency of emissions constraints and technical performance of LDES. Finally, in Section \ref{sec:conclusions} we also draw conclusions both from this study and from other recent work to present a guide for modelers establishing best practices for modeling LDES in capacity expansion models.

It is important to clarify that, since LDES is not a term with a universally accepted definition, in this study we take LDES to mean energy storage that is capable of discharging for multiple days at its full capacity (sometimes also referred to as multi-day energy storage). It is these resources in particular that many existing energy system models are not able to accurately represent and are therefore the focus of this study.

\section{Material and methods}
\label{sec:methods}

\subsection{Experimental design}

The core experimental design pursued by our study is to run our capacity expansion model for varying amounts of spatial and temporal domain reduction (or abstraction), and examine how the estimated marginal value of LDES capacity changes in each case. In all cases, we run capacity expansion cases for the entire continental United States for 2045. In most cases, we incorporate a zero carbon emissions constraint, as we wish to study the response of LDES value to model abstraction when the value is highest; however, some of our sensitivity cases explore the impact of relaxing this emissions constraint. 

To estimate the value of LDES in each case, we force in a small amount of zero-cost, 200-hr, 42\% round-trip efficiency LDES with a fixed energy storage capacity or duration, and examine the shadow price of the constraint that is used to force in the resource. This shadow price, also referred to as dual value or Lagrange multiplier, denotes the marginal change in the objective function of our model with regard to a marginal change in the right hand side of the constraint. Since in this case, our objective function is the total electricity system cost, and the right hand side of the constraint is the maximum capacity in MW of LDES capacity, the shadow price denotes the marginal reduction in total system cost per MW of LDES capacity. This system value can also be interpreted as the maximum  `break-even' annualized capacity cost of LDES such that the marginal reduction in system costs is greater than the cost of LDES capacity (for the amount of capacity forced in) and thus LDES capacity can be considered economically justified. This basic approach to estimating marginal value of resources has been used previously in several recent studies \cite{mallapragada_long-run_2020, schwartz_value_2023}. 

When we vary model temporal resolution, we examine both varying numbers of representative periods, and varying representative period lengths. We examine a range of between 5 and 100 representative periods, and a range of representative period lengths from 24 hours to 2 weeks. In addition to these representative period experiments, we also run a full year ``monolithic'' case consisting of 8760 hours. When we vary model spatial resolution, we examine a range of spatial aggregations for the continental United States of between 3 and 26 zones.

In addition to this core experiment, we also examine the robustness of our findings to several key dimensions of uncertainty. These dimensions are: a) the cost of other firm zero carbon resources, b) the availability of other firm zero carbon resources, c) the amount of LDES forced in, d) the duration of LDES forced in, e) the emissions constraint, f) the round-trip efficiency, and g) the inclusion of the ability for LDES to be co-located with renewables. We chose these parameters based on our understanding of the primary drivers of LDES value, as determined in previous work such as \cite{sepulveda_design_2021, dowling_role_2020}.

Finally, we show the impact of the period-linking constraints and capacity reserve margin formulation (two key parts of our modeling framework) on our results, and also demonstrate the importance of using the best practices identified by this work in the context of accurately capturing the costs of deep decarbonization.

\subsection{Modeling framework details}
\label{sec:modeling-details}

We use the open-source GenX capacity expansion model \cite{jenkins_enhanced_2017}, which is a flexible and configurable framework for building capacity expansion optimization models based on the Julia programming language \cite{bezanson_julia_2017} and JuMP package for mathematical programming \cite{lubin_jump_2022}. In short, GenX seeks to optimize the least-cost portfolio of electricity generation, storage and inter-zonal transmission capacity to meet electricity demand for a given system, and it incorporates inter-zonal power flow constraints and detailed operational constraints at hourly resolution for a set of representative operational periods. The optimization problem GenX creates is usually linear (as in our cases), but it can be mixed integer if full unit commitment and/or integer capacity decisions are employed. The details of GenX's formulation as they relate to the current study are described in further detail in Section \ref{sec:formulation} and the source code is available on GitHub \footnote{Available at: \url{https://github.com/GenXProject/GenX.jl}}.

To create inputs for GenX, we use the open-source PowerGenome tool \cite{schivley_powergenomepowergenome_2022}, which creates capacity expansion inputs based on publicly available energy system data. The primary purposes of this tool are to aggregate operational and cost parameters of all existing and new generators, develop network constraint data, and develop spatially aggregated clusters of candidate variable renewables (wind and solar PV) along with their associated availability profiles and estimated transmission interconnection costs. When PowerGenome determines the interconnection cost for new variable renewable resources, it uses an algorithm that determines this cost based on not only the cost of a spur line to the transmission system, but also on the cost of delivery to demand centers (including transmission network upgrades), which is dynamic based on the zonal aggregation being used. This is a novel technique, described further in \cite{patankar_land_2023}, which we expect reduces the error associated with spatial domain reduction (due to a more accurate estimation of the cost of delivery to demand within a modeled zone even in more spatially aggregated systems), although we have not yet conclusively demonstrated how well this holds under various levels of spatial aggregations. PowerGenome also compiles data for electricity demand projections, which it obtains from the NREL Electrification Futures Study \cite{mai_electrification_2018}, and for existing generators, which it obtains from EIA Form 860 \cite{us_energy_information_administration_eia_2023} along with other EIA sources.

In all cases, we use resource performance and cost inputs from the National Renewable Energy Laboratory's Annual Technology Baseline (ATB) report, 2022 edition \cite{nrel_2022_2022}. All renewable and fossil technologies in the ATB are assumed to be available, including advanced nuclear and natural gas CCGT with CCS. We also add a ``Zero Carbon CT'' resource which is identical to the Natural Gas CT (combustion turbine) in the ATB, but with a \$20/MMBTU fuel cost, intended to represent a generic zero carbon fuel, such as renewable natural gas, synthetic methane, or hydrogen. We chose this cost as the future availability and cost of zero carbon fuels is highly uncertain, and because this value aligns with some prior work on the potential cost of renewable natural gas \cite{icf_renewable_2019}. As the cost and availability of nearly all zero carbon firm technologies is highly uncertain, we focus on these parameters in the uncertainty analysis section of this study. All costs used in the study are presented in real 2023 US dollars (2023\$) and as a result, the estimated LDES values reported herein are in 2023\$ as well.

\subsection{Capacity expansion model details}
\label{sec:formulation}

The GenX capacity expansion model is described in depth in other publications such as \cite{jenkins_enhanced_2017} and the model documentation\footnote{Available at: \url{https://github.com/GenXProject/GenX.jl}}, and we will not repeat the full formulation here. Instead, we briefly highlight some aspects of the modeling framework and formulation that are relevant to the current study.

\subsubsection{Time domain reduction}

GenX is built to work with either representative periods or a full year of operational data at hourly resolution, and it contains functionality to perform time domain reduction to develop the representative periods. The time domain reduction is performed by first including the ``extreme'' system periods of minimum solar, minimum wind, and maximum load as representative periods, and by subsequently using a k-means clustering algorithm to cluster the rest of the operational year into representative periods \footnote{See description at: \urltdr}.

\subsubsection{Long duration energy storage formulation}

\begin{figure}[htbp]
  \centering
  \includegraphics[width=1\textwidth]{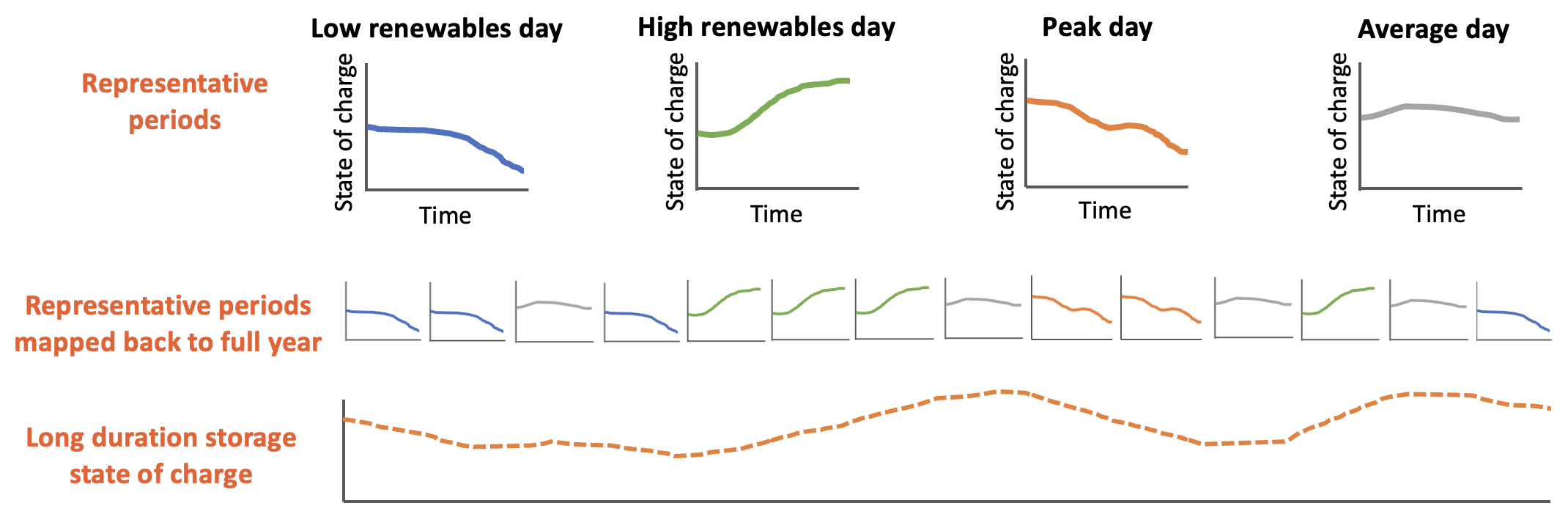}
  \caption{Illustration of LDES period-linking methodology developed by Kotzur et al. \cite{kotzur_time_2018} for four representative days. The top row shows LDES state of charge for each representative period. Operations of the full system are modeled for these representative periods at hourly resolution. The middle row shows the sequence that the representative periods are arranged in to develop an approximated version of the full year (in reality this would have 365 periods, but only a few are shown for simplicity). The bottom row shows LDES state of charge across this approximated representation of the full year, which is calculated using the change in state of charge for each period in the sequence in the middle row. Note that this formulation allows LDES to both shift energy between seasons, and to discharge for longer than the length of a representative period, neither of which would be possible if the periods were unlinked.}
  \label{fig:diagram}
\end{figure}

To model long duration energy storage resources, GenX implements the period-linking methodology developed by Kotzur et al. \cite{kotzur_time_2018}, which allows the multi-day and seasonal shifting ability of LDES to be captured, while still using representative periods to model system operations. In brief, the algorithm works by tracking the state of charge of LDES across an approximated version of a full year, which is developed by composing a full year out of only the representative periods. Figure \ref{fig:diagram} shows an illustration of this methodology for an illustrative system with only four representative days.

The more technical description of this algorithm is that it models LDES state of charge by modeling both an intra-period and inter-period component, where the former describes changes in state of charge and operations during a representative period (at hourly resolution) and the latter tracks changes in state of charge and ensures feasibility across representative periods (at the resolution of one representative period length, however defined). This is accomplished by maintaining the standard state of charge tracking for LDES resources within a representative period (the intra-period component), while tracking the cumulative increase or decrease in state of charge across each representative period,  (the inter-period component). Thus,  in this period-linking formulation, the constraint which couples the state of charge for storage at the end of a representative period with that of the beginning of the same period (the ``state of charge wrapping constraint'') becomes:

\begin{equation}
\begin{aligned}
& \Gamma_{o,z,(m-1)\times \tau^{period}+1 } =\left(1-\eta_{o,z}^{loss}\right)\times \left(\Gamma_{o,z,m\times \tau^{period}} -\Delta Q_{o,z,m}\right) -  \\
& \frac{1}{\eta_{o,z}^{discharge}}\Theta_{o,z,(m-1)\times \tau^{period}+1} + \eta_{o,z}^{charge}\Pi_{o,z,(m-1)\times \tau^{period}+1} \quad \forall o \in \mathcal{O}^{LDES}, z \in \mathcal{Z}, m \in \mathcal{M}
\end{aligned}
\end{equation}

Where $\Gamma_{o,z,t}$ is the intra-period state of charge of the storage resource, $\eta_{o,z}^{loss}$ is the self discharge or parasitic loss rate, $\Delta Q_{o,z,m}$ is the change in inter-period state of charge of the storage resource for the current representative period, $\eta_{o,z}^{discharge}$ and $\eta_{o,z}^{charge}$ are the discharging and charging efficiencies, $\Theta_{o,z,t}$ and $\Pi_{o,z,t}$ are the discharging and charging of the resource for a time step $t$ within a representative period (e.g. hour). $\mathcal{O}^{LDES}$ is the set of all storage resources, $\mathcal{Z}$ is the set of zones, and $\mathcal{M}$ is the set of representative periods. Finally $\tau^{period}$ is the number of time steps per representative period. Note that $(m-1)\times \tau^{period}+1$ refers to the first time step of representative period $m$, and $m\times \tau^{period}$ refers to the last time step of the same period. The state of charge $\Gamma_{o,z,t}$ for a time step is the state of charge at the end of a time step.

In addition to this modification, the inter-period state of charge is separately tracked in a new set of constraints, which take as an input the function mapping $f: n \rightarrow m$ which uniquely maps each period of the full operational year $n$, which we refer to as an ``input period,'' to the corresponding representative period $m$ modeled at hourly resolution. This mapping comes from the representative period selection process, which chooses a set of representative periods from the set of input periods (for example, 10 representative days might be chosen from the set of 365 available days), and includes the mapping of input periods to representative periods as an output. First, we link the inter-period state of charge  $Q_{o,z,n}$ for each input period $n$ to the change in inter-period state of charge $\Delta Q_{o,z,m}$ for the corresponding representative period (this is equivalent to setting $Q_{o,z,n}$ to the true state of charge, accounting for the $\Delta Q$, at the end of the corresponding representative period):

\begin{equation}
\begin{aligned}
& Q_{o,z,n} =\Gamma_{o,z,f(n)\times \tau^{period}} - \Delta Q_{o,z,m}
\quad \forall  o \in \mathcal{O}^{LDES}, z \in \mathcal{Z}, n \in   \mathcal{N}
\end{aligned}
\end{equation}

Where $\mathcal{N}$ is the set of input periods.

Then, the inter-period state of charge $Q_{o,z,n}$ is tracked over the course of the full year (the full set of input periods):

\begin{equation}
\begin{aligned}
& Q_{o,z,n+1} = Q_{o,z,n} + \Delta Q_{o,z,f(n)}
\quad \forall  o \in \mathcal{O}^{LDES}, z \in \mathcal{Z}, n \in \mathcal{N}\setminus\{|N|\}
\end{aligned}
\end{equation}
\begin{equation}
\begin{aligned}
& Q_{o,z,1} = Q_{o,z,|N|} + \Delta Q_{o,z,f(|N|)}
\quad \forall  o \in \mathcal{O}^{LDES}, z \in \mathcal{Z}, n = |N|
\end{aligned}
\end{equation}

Note that $|N|$ refers to the last input period of the year.

Finally, we constrain the inter-period state of charge to not exceed the storage capacity of the resource $\Delta^{total, energy}_{o,z}$:

\begin{equation}
\begin{aligned}
    Q_{o,z,n} \leq \Delta^{total, energy}_{o,z}
\quad \forall n \in \mathcal{N}, o \in \mathcal{O}^{LDES}
\end{aligned}
\end{equation}

It should be emphasized that this formulation is functionally nearly identical to the formulation described in Kotzur et al. \cite{kotzur_time_2018}, except that our formulation tracks the ``true'' or total state of charge in both the intra-period and inter-period components, whereas in Kotzur et al. the authors decompose the state of charge into two components, which are then added to obtain the ``true'' state of charge. We find that the formulation in Kotzur et al. can permit state of charge to take on physically infeasible values during a representative period if there is non-monotonic charging/discharging behavior during the period, while our subtle improvement in the formulation prevents most (but not all) of these infeasible outcomes. It is also important to note that we use this same formulation for reservoir hydro resources in GenX when using representative periods.

\subsubsection{Capacity reserve margin formulation}
\label{sec:capres}

In the version of GenX used in this study, we use a novel formulation for determining the contribution of storage resources towards the capacity reserve margin (CRM, also known as planning reserve margin or PRM), which we included in an effort to make sure LDES is not disadvantaged relative to other clean firm resources in our modeling. This formulation is described further in \ref{sec:prm} but summarized briefly here as it is a distinct methodology from previous studies using GenX.

The capacity reserve margin constraint, which ensures that a surplus amount of capacity is built on top of peak demand, is included to ensure that the portfolio selected by GenX has an acceptable level of reliability across all weather conditions and expected contingencies (e.g. generator or transmission outages), not just those in the modeled weather year. This constraint requires an estimation of each resource's effective capacity contribution, that is, the capacity that is counted towards the capacity reserve margin, which is less than or equal to the actual nameplate capacity of the resource. For firm resources such as gas turbines, this capacity contribution is set equal to 95\% of the resource's nameplate capacity; the de-rating is included to account for forced outages. For energy-limited resources such as VRE, hydro, and storage, this capacity contribution is calculated endogenously by counting only the amount that each resource is actually available to generate during peak hours; this contribution is further de-rated by 20\% to account for inter-annual variability. This de-rate value should ideally be informed by multi-year weather data and established independently for each resource class, but here we select a reasonable illustrative value. This method of endogenously calculating capacity value is implemented to ensure that the interactive effects of resource capacity value are captured and to avoid having to run a separate resource adequacy model to determine capacity value.

For storage resources, it is not trivial to determine a resource's availability to generate during peak hours, which is needed to determine its capacity contribution. If a storage resource is generating at full output during the peak hour in the model, then it is clear that it has the availability to generate during that hour, as the resource has charged during other hours to enable this behavior. However, if it is not generating at full output, then it is not clear whether it has the ability to increase its output if called upon during some contingency, as this would require charging more in previous hours, and/or reducing discharge in subsequent hours. Therefore, it is not clear to what extent storage should contribute to the capacity reserve margin constraint on top of its actual generation during peak hours. We solve this problem by developing a formulation that accounts for the capacity contribution of storage beyond its actual generation in peak hours, taking into account behavior in other hours. In brief, storage is given an ability to ``virtually'' discharge, which is backed up by a ``virtual state of charge'' which represents energy capacity that is set aside just for the virtual discharge. This formulation ensures that a storage resource holding state of charge `in reserve' can contribute to capacity reserve requirements, while ensuring that the physical capacity constraints of the storage resource are respected in both actual modeled dispatch and hypothetical contributions to reserves during contingencies. We describe this novel formulation in more detail in \ref{sec:prm}.

\section{Results}

In this section, we first examine the impact of temporal resolution on LDES value and examine the causes of the trends we find. Second, we investigate the impact of representative period length on LDES value. Third, we examine the impact of spatial resolution. Fourth, we examine the impact of varying fixed costs of clean firm resources and varying LDES durations, and briefly discuss the impact of other key dimensions of uncertainty. Finally, we examine the impact that the period-linking methodology has on our results.

\subsection{Impact of temporal resolution on LDES value}

\begin{figure}[tbp]
  \centering
  \includegraphics[width=1\textwidth]{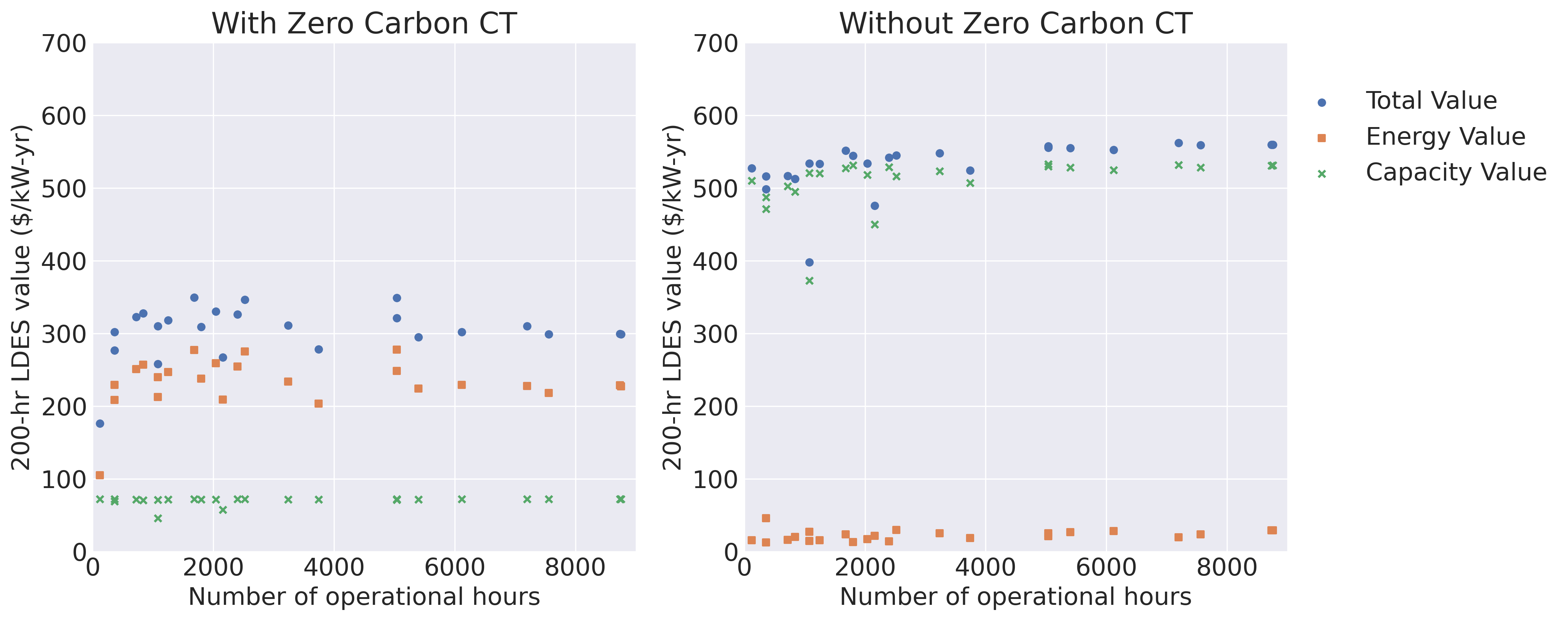}
  \caption{LDES value as a function of number of operational hours modeled, with and without the Zero Carbon CT resource, and with a breakout of the total value into capacity and energy value. Each dot represents one run of GenX with a particular number and length of representative periods.}
  \label{fig:main}
\end{figure}

\begin{figure}[tp]
  \centering
  \includegraphics[width=1\textwidth]{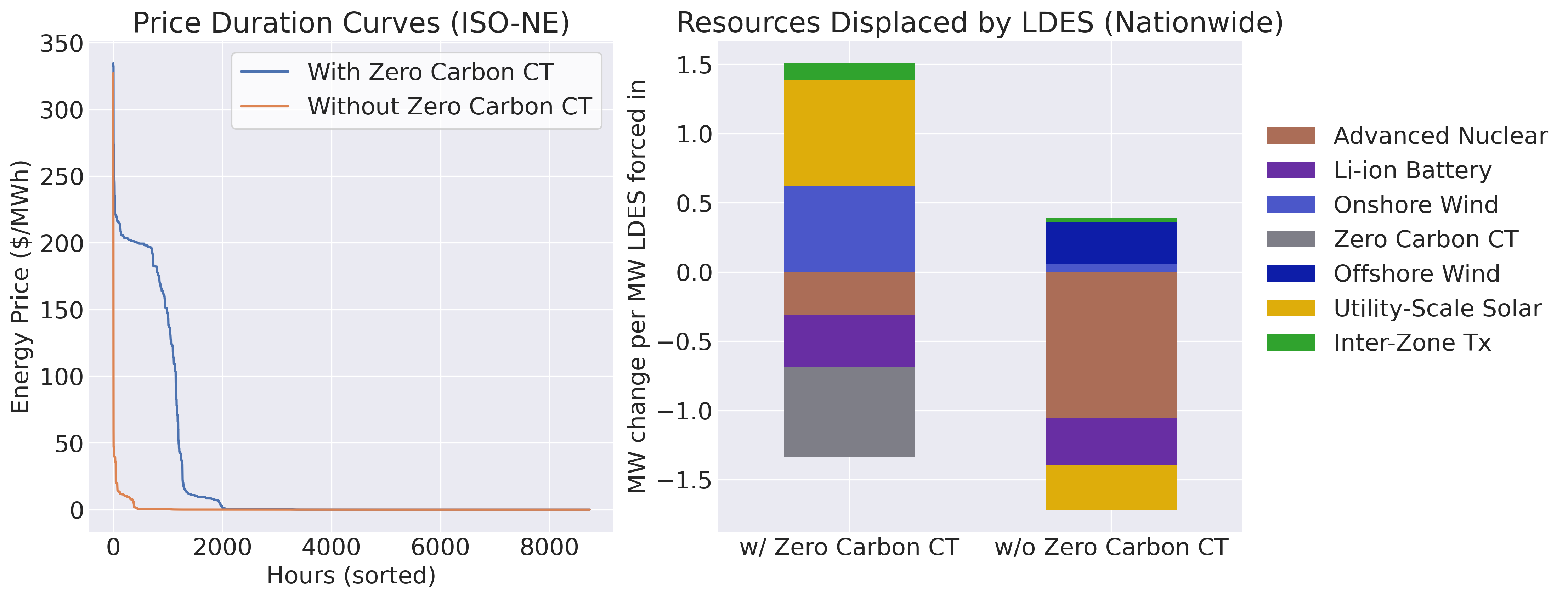}
  \caption{Left: price duration curves for the ERCOT zone with and without the Zero Carbon CT resource. Right: change in other resources when LDES is forced in (summed across all zones), with and without the Zero Carbon CT resource.}
  \label{fig:price_duration_and_resources_displaced}
\end{figure}

Figure \ref{fig:main} shows 200-hr LDES value, broken out into energy value, capacity value, and total value, as a function of number of operational hours modeled, for two main cases: with the Zero Carbon CT resource included and without this resource included. These two scenarios are included because the results differ significantly with and without this resource. We will begin with describing the sources of LDES value in each case, and then move to the implications for trends in value as a function of temporal resolution. When the Zero Carbon CT resource is included (left), LDES derives its value mainly from energy arbitrage (orange squares), and to a lesser extent from displacing capacity resources (green x's). This happens because the Zero Carbon CTs, which serve as a capacity resource in this case, are relatively inexpensive capacity but are very expensive to run due to high fuel costs. As a result, they generate infrequently but result in very high prices when they do generate, increasing the volatility of electricity prices and enabling storage devices to earn significant revenue from energy price arbitrage. These price dynamics can also be seen in Figure \ref{fig:price_duration_and_resources_displaced}, which shows the price duration curve for one zone in the model, with and without the Zero Carbon CT resource included, as well as the change in installed capacity of other resources for each MW of LDES capacity deployed. LDES does derive capacity value from displacing these resources, but this value is relatively low since the capital cost of Zero Carbon CTs is on the order of \$70/kW-yr. On the other hand, when the Zero Carbon CT resource is not included (right side of Figure \ref{fig:main}), LDES value comes mainly from capacity value (green x's), and much less so from energy arbitrage value (orange squares). In these cases, Advanced Nuclear is serving as the primary capacity resource to meet (net) peak demand and satisfy the capacity reserve margin constraint. This resource has a much higher capital cost, on the order of \$500/kW-yr, and since LDES can displace it, the capacity value of LDES is much higher in this case. However, since advanced nuclear are much cheaper to run than the Zero Carbon CTs, nuclear generates more often during the year and consequently sets much lower prices for much of the year (Figure \ref{fig:price_duration_and_resources_displaced}), which reduces the opportunity for LDES to earn energy revenue via arbitrage. In summary, if a low capital cost / high variable cost resource like the Zero Carbon CT is the marginal firm resource contributing to capacity reserve requirements, LDES is likely to derive its value mainly from energy arbitrage and slightly from capacity displacement. In contrast, when a high capital cost / low variable cost resource like Advanced Nuclear (alternatively: geothermal) is the marginal capacity resource, LDES derives its value mainly from capacity displacement. 

These dynamics have important implications for the trend in LDES value as a function of modeled temporal resolution. Figure \ref{fig:main} shows that not much temporal resolution is necessary to get an accurate picture of capacity value for LDES (green x's), whereas for energy value, convergence towards the ``true'' value is only achieved starting around 6000 operational hours. This means that, for cases where LDES is primarily providing capacity value, like where it is primarily substituting for a high fixed cost, low variable cost resource like Advanced Nuclear (the ``Without Zero Carbon CT case''), not much temporal resolution is needed to get an accurate picture of LDES value. For cases where LDES is both providing capacity value and also derives a significant amount of value from energy arbitrage, (the ``With Zero Carbon CT'' case), much more temporal resolution is required to converge to the ``true'' value. In these cases, it is also likely that the value would not be stable across multiple modeled weather years, as the interannual variability of renewables and demand can be significant, causing variation in the frequency and duration of periods when high variable cost resources like CTs are operating and set high marginal prices. These results mean that how much temporal resolution is necessary to get an accurate picture of LDES value will be dependent on the specifics of the case being modeled and the composition of the resource portfolio deployed alongside LDES.


\subsection{Impact of representative period length on LDES value}

\begin{figure}[htbp]
  \centering
  \includegraphics[width=1\textwidth]{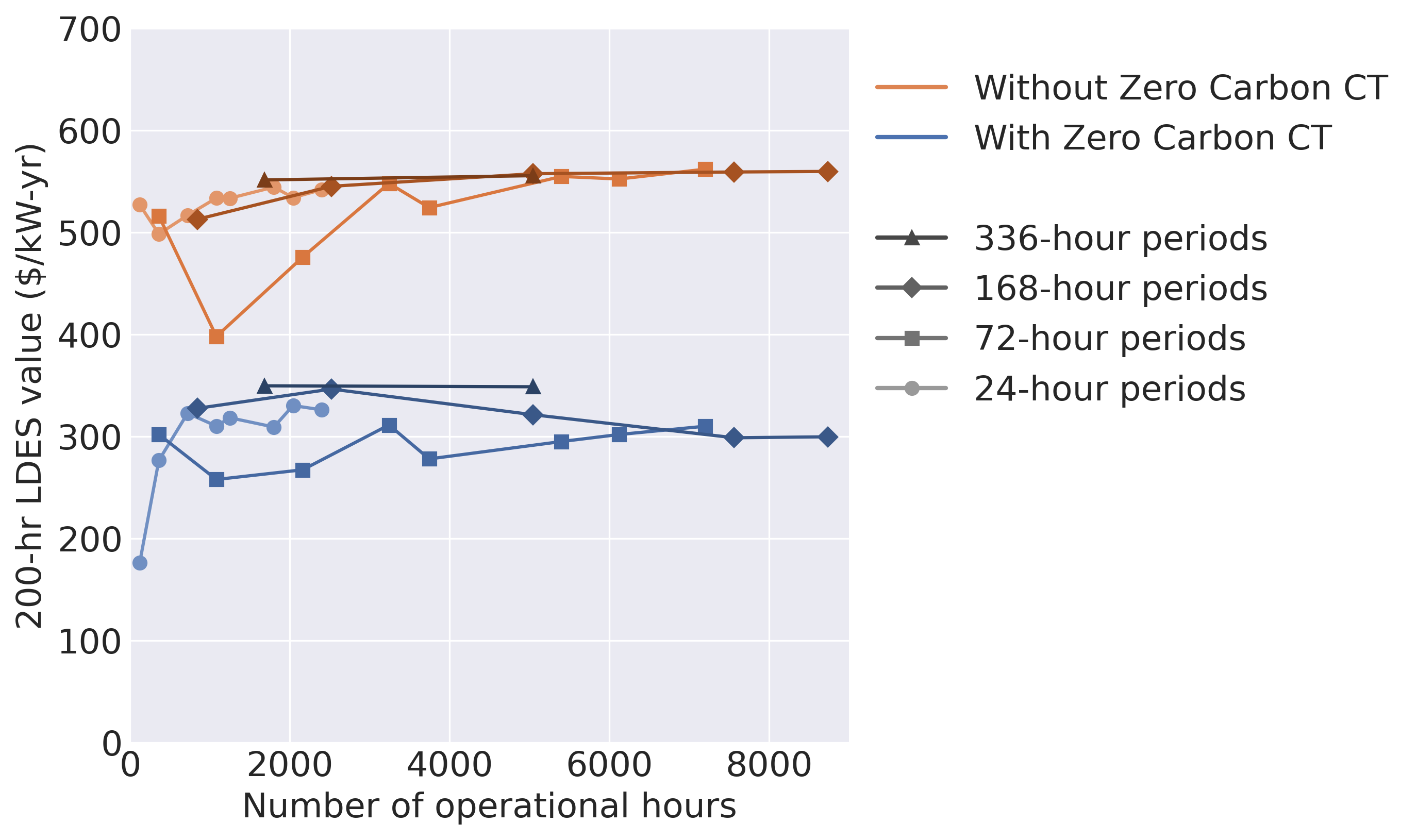}
  \caption{LDES value as a function of number of operational hours modeled, broken out by representative period length.}
  \label{fig:hours_breakout}
\end{figure}

Figure \ref{fig:hours_breakout} shows the results of the same experiment as is shown in Figure \ref{fig:main}, but broken out by representative period length. Generally, there is a trend of increasing LDES value with increasing representative period length (particularly at low temporal resolutions), but this trend is not always present, and in some cases the longer periods can over-represent LDES value (darkest blue line). This is likely because there is error introduced by the constraint of having to pick only 5 or 15 two week periods in the time domain reduction algorithm- relative to being able to pick 10 or 30 week-long periods, which would have the same number of operational hours but provide more degrees of freedom to better approximate renewables availability time series throughout the course of the year. Therefore, picking longer representative periods is not always better, particularly when restricted to a limited number of operational hours overall, because selecting only a handful of longer representative periods could lead to a less accurate picture of the distribution of renewables availability over the course of the year.

\subsection{Impact of spatial resolution on LDES value}

\begin{figure}[htbp]
  \centering
  \includegraphics[width=1\textwidth]{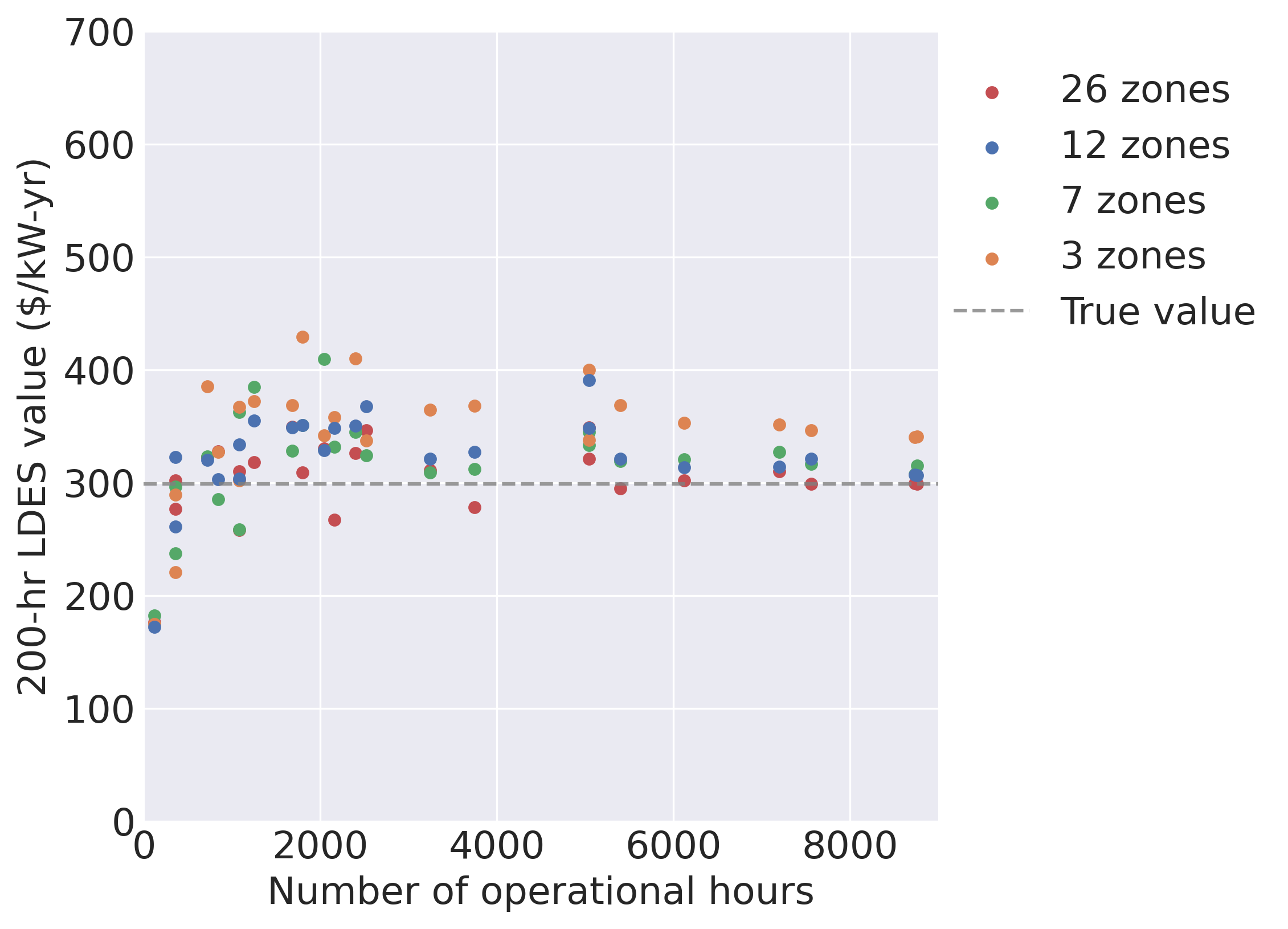}
  \caption{LDES value as a function of number of operational hours included, for multiple zonal aggregations. The two orange dots present for 8760 hours are a result of running both a full year without time domain reduction, and 52 representative weeks, which are not identical because in the latter case only LDES is able to shift energy between weeks. 8760 results are not included for all zones because the 8760 model was not computationally feasible for more granular zonal aggregations.}
  \label{fig:impact_of_zones_all}
\end{figure}

Figure \ref{fig:impact_of_zones_all} shows the impact of spatial resolution on LDES value. The results indicate that, in our experiment, spatial resolution does not have as significant of an effect on LDES value as does temporal resolution. Indeed, after a sufficiently high temporal resolution, the difference in estimated LDES value across different spatial resolutions is fairly small (less than +/-8\% for all cases with $>$6000 hours). We hypothesize that this result is due to our interconnection cost methodology (described in Section \ref{sec:modeling-details}), which adjusts interconnection costs to account for delivery to load as the spatial resolution changes; however, we did not test this hypothesis. Therefore, LDES value may be more sensitive to spatial resolution if a different interconnection cost methodology is used.

It is important to note that, while we find that LDES value is not significantly impacted by spatial resolution, other studies have found that other outcomes of potential interest, such as total cost, capacity, and renewables siting, are more sensitive to spatial resolution and would likely exhibit decreased error under higher spatial resolution \cite{martinez-gordon_review_2021, krishnan_evaluating_2016, frysztacki_inverse_2023, frysztacki_strong_2021}. 

\subsection{Sensitivity of LDES value to changes in costs of other resources, and storage duration}

\begin{figure}[htbp]
  \centering
  \includegraphics[width=1\textwidth]{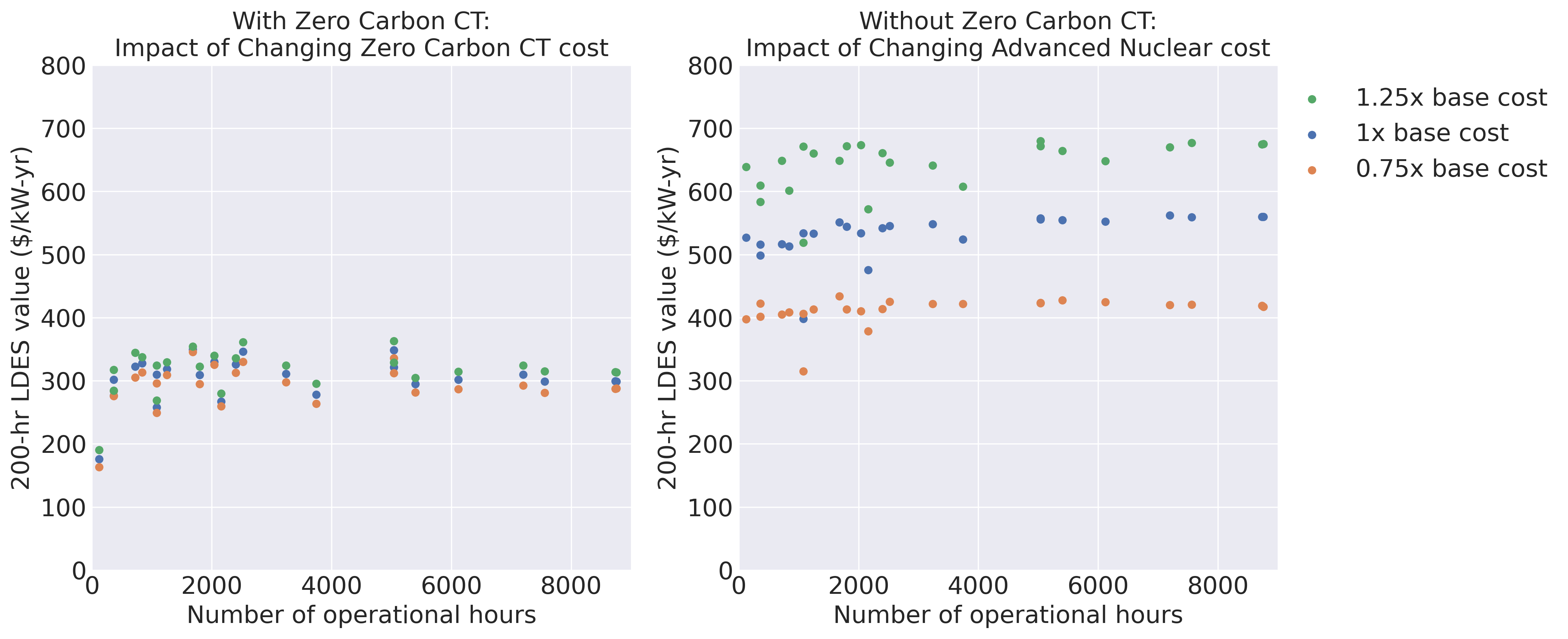}
  \caption{Left: LDES value as a function of number of operational hours modeled for the case with Zero Carbon CT included, with the fixed cost varied up and down by a factor of 25\%. Right: Same experiment, but for the case without the Zero Carbon CT included, and with the fixed cost of the Advanced Nuclear resource varied.}
  \label{fig:cost_sensitivity}
\end{figure}

Figure \ref{fig:cost_sensitivity} shows the sensitivity of the main two experiments (with and without the Zero Carbon CT) to changes in the fixed cost of the Zero Carbon CT resource (left, for the case where this resource is included), and to changes in the fixed cost of Advanced Nuclear (right, for the case where the Zero Carbon CT resource is not included). It is clear from these results that: a) the value of LDES is highly dependent on the cost of resources it is substituting for, and b) the results in terms of convergence trends are not sensitive to these changes in fixed cost.

\begin{figure}[htbp]
  \centering
  \includegraphics[width=1\textwidth]{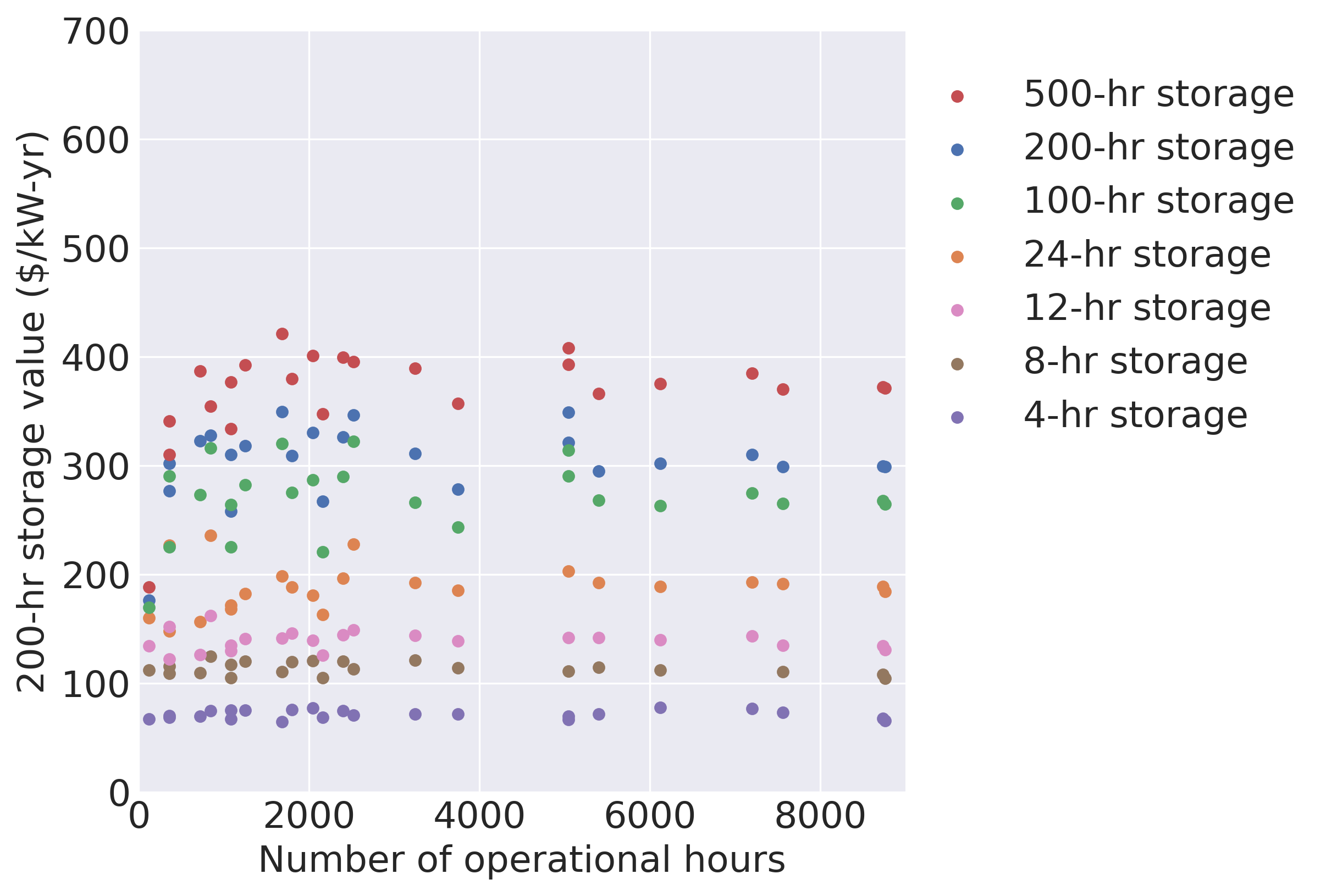}
  \caption{LDES value as a function of number of operational hours modeled, but for varying durations down to 4 hours. All experiments in this figure have a constant round trip efficiency for the forced-in storage resource of 42\%.}
  \label{fig:impact_of_duration}
\end{figure}

Figure \ref{fig:impact_of_duration} shows the sensitivity of the main case with the Zero Carbon CT resource included to changes in the duration of the storage device forced in, from 500 hours down to 4 hours (these results are shown for a constant round trip efficiency of 42\%). These results show that, for storage of 12 hours and shorter in duration, temporal resolution is much less important for capturing the full value of the device. Therefore, the results of this study mainly apply to storage of a duration longer than 24 hours (multi-day storage).

Figures \ref{fig:impact_of_amount_forced_in}, \ref{fig:emissions_impact}, \ref{fig:rte}, \ref{fig:colocation_impact}, \ref{fig:zone_trends} in \ref{sec:extra_results} additionally show that the key results of this study are not sensitive to: a) the amount of LDES forced in, b) the presence of a less-tight emissions constraint, c) the round-trip efficiency of the LDES, d) the ability of LDES to be co-located with renewables, or e) the interconnection that LDES is located in (since the three US interconnections have quite different resource mixes, this is an indication that our results are generalizable to other systems). Additionally, Figure \ref{fig:impact_of_no_nuclear_12z} shows that, when no other clean firm resources are available (meaning the system is relying only on renewables, Li-ion batteries, and LDES to meet load), LDES value becomes even more noisy and does not show a clear convergence trend. However, it has been well documented that achieving absolute zero emissions without any clean firm resources or long duration storage is not economically feasible or realistic \cite{jenkins_getting_2018}, so we opt not to focus on this case.

\subsection{Impact of period-linking constraints on results}

\begin{figure}[htbp]
  \centering
  \includegraphics[width=1\textwidth]{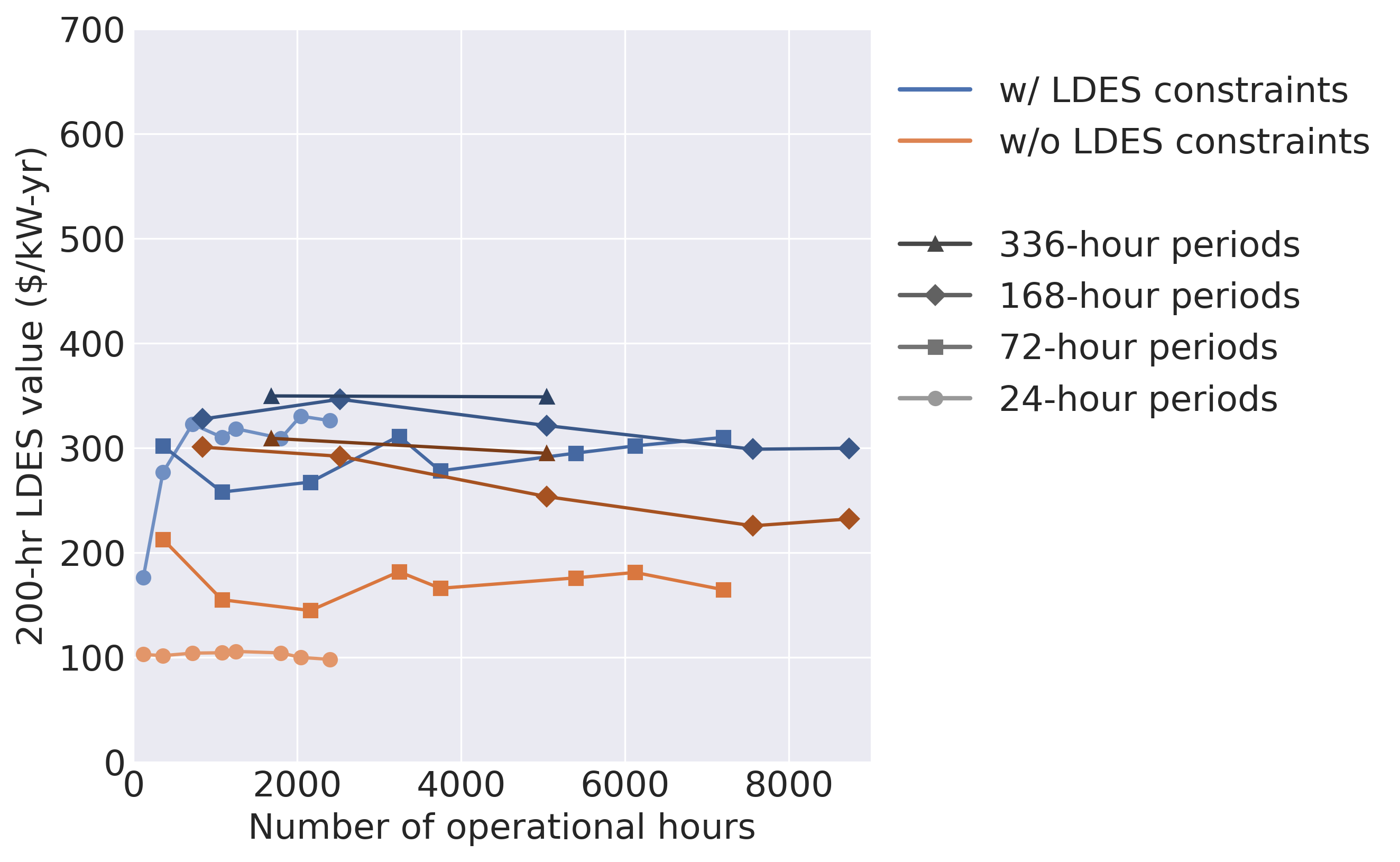}
  \caption{LDES value as a function of number of operational hours included, with (blue) and without (orange) the LDES period linking constraints. Results are broken out to show the results for different representative period lengths, as shown in the legend.}
  \label{fig:impact_of_no_ldes_constraints}
\end{figure}

For the sake of modelers wishing to know how important the period-linking methodology is to the conclusions reached in this study, here we show our results with and without the period-linking methodology turned on (Figure \ref{fig:impact_of_no_ldes_constraints}). These results indicate that the period-linking methodology is crucial for accurately capturing the value of LDES when representative periods are used, even when using a very high number of operational hours. Even 52 unlinked representative weeks (second orange line from top) are insufficient to capture the full value of LDES, as this situation does not allow LDES to shift energy between weeks. Of particular note is the bottom orange line which represents unlinked representative days, varying in number from 5 to 100. These unlinked representative days, which are commonly seen as a best practice, lead to an underestimation of the value of LDES by around 70\%, even when 100 representative days are included. 

Relatedly, in Figure \ref{fig:with_without_vdis} in Section \ref{sec:extra_results}, we demonstrate that the capacity reserve margin formulation with ``virtual discharge'' described in this paper is a less important factor in capturing the value of LDES.

\begin{figure}[htbp]
  \centering
  \includegraphics[width=1\textwidth]{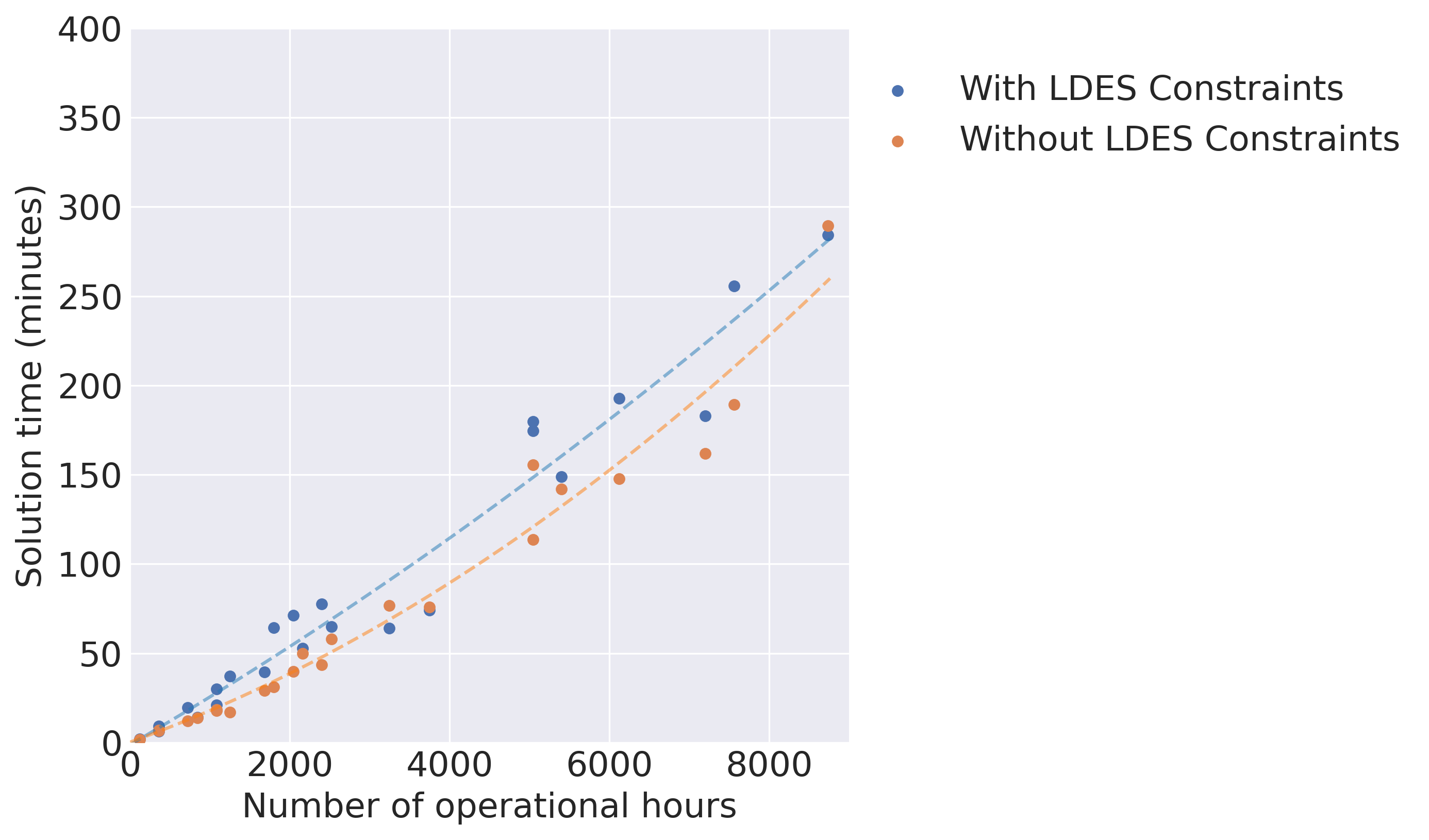}
  \caption{Model solution time as a function of number of operational hours included, with and without LDES period linking constraints.}
  \label{fig:solution_time_withandwithoutldes}
\end{figure}

We also present the solution time of our model runs as a function of number of operational hours, to emphasize that the LDES period-linking methodology can approximate the full value of LDES without modeling a full year with reasonable accuracy while delivering significant savings in terms of solution time. Figure \ref{fig:solution_time_withandwithoutldes} shows that solution time scales roughly quadratically with the number of operational hours, and that solution time is generally slightly increased when the LDES period linking constraints are included, although there is not a clear trend. Thus, the LDES constraints do increase computational complexity (by way of adding more constraints that couple the periods), but they do not appear to have a significant impact on runtime (for an equivalent number of total time periods modeled) in most of the cases modeled. Thus, using representative periods with the LDES period linking constraints allows for a significant savings in runtime compared to modeling a full year of operations while still ensuring that LDES value is accurately captured, and perhaps more importantly, it allows for modeling of cases that would not even be computationally feasible (e.g., too big to fit in memory) if modeling a full year.

\subsection{Impact of LDES modeling best practices on accurately capturing cost of decarbonization}

\begin{figure}[htbp]
  \centering
  \includegraphics[width=1\textwidth]{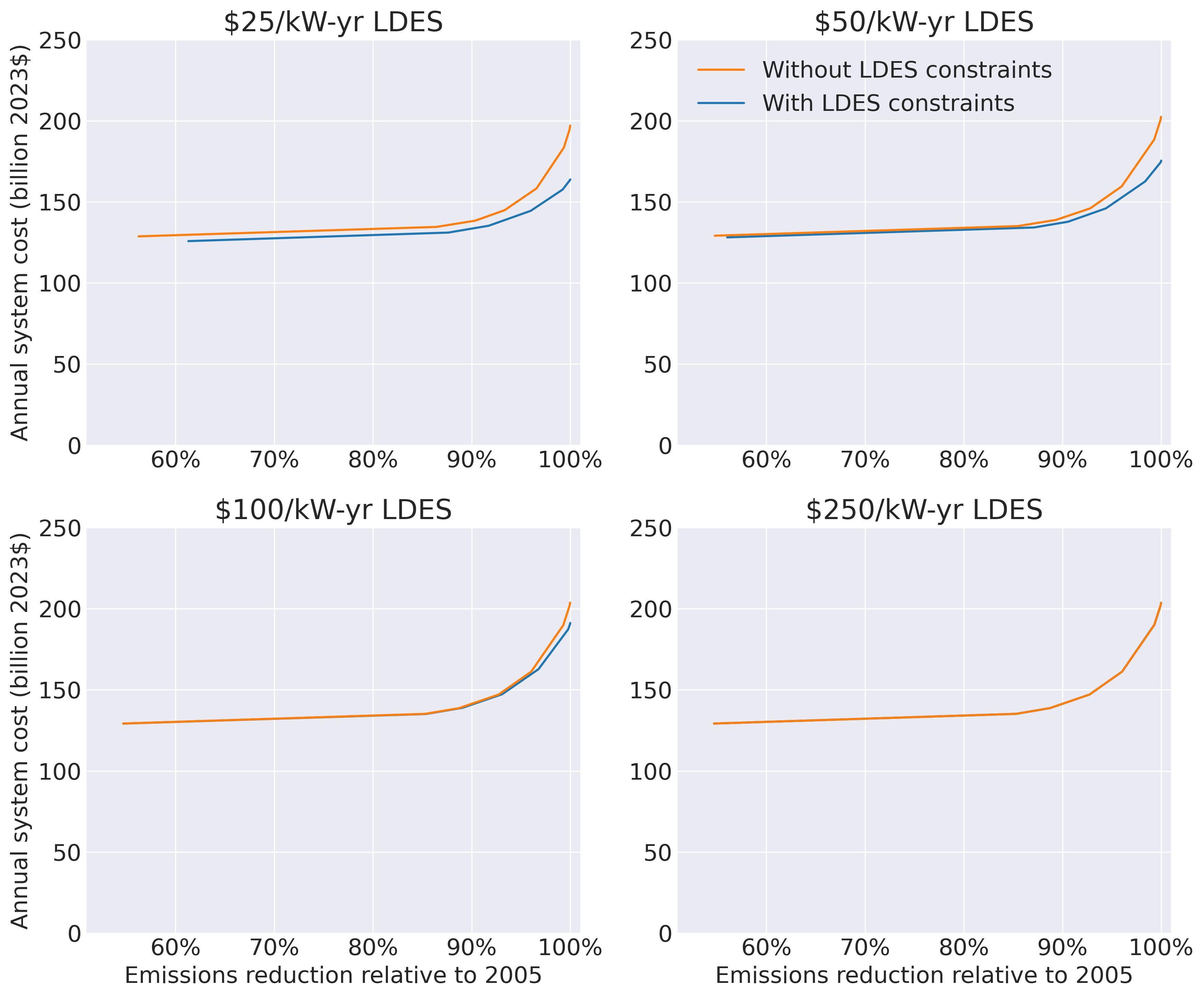}
  \caption{Total annual (US-wide) electricity system cost as a function of emissions constraint, with and without LDES constraints. Model runs are performed using 45 representative days. Emissions reduction is induced using a carbon price ranging from \$25/ton to \$2000/ton, and not using a hard cap.}
  \label{fig:ldes_decarb_cost}
\end{figure}

In this final results section, we investigate the impact of the LDES modeling best practices described on this paper on a key output from capacity expansion models, which is the cost of varying levels of decarbonization. Figure \ref{fig:ldes_decarb_cost} shows that, for very cheap LDES in particular, not including the LDES period-linking constraints can cause a significant overestimation of the costs of deep decarbonization. Specifically, the cost of near-complete (99.99\%) decarbonization is overestimated by 20\% for \$25/kW-yr 200-hr LDES, or 15\% for \$50/kW-yr 200-hr LDES, if the LDES period-linking constraints are not included. The impact of LDES constraints on accurately capturing the cost of decarbonization is minimal for 200-hour LDES costing more than \$100/kW-yr. These results demonstrate that using the best practices described in this paper, and in particular, using the period-linking constraints, is crucial for accurately capturing the cost of deep decarbonization in capacity expansion models. They also demonstrate, as other studies have shown, that LDES must be very inexpensive to have a significant impact on the cost of decarbonization \cite{sepulveda_design_2021, cole_cost_2023}.

\section{Discussion}

Two key overall findings can be drawn from our results. 

Firstly, we find that the value of LDES is highly sensitive to the modeled temporal resolution, \textit{and} to how representative operational periods are linked. How much temporal resolution is needed to capture the full value of LDES will depend on the specific case being modeled. If LDES serves mainly to displace capital-intensive, low-variable cost firm capacity resources (e.g., nuclear, geothermal) and energy arbitrage opportunities are limited, then the full value can likely be represented with relatively few operational hours (e.g. on the order of 2500). If LDES derives a significant amount of value both from displacing firm capacity resources and from energy price arbitrage (as in a case with high variable cost firm resources present like Zero Carbon CTs), then much more temporal resolution is needed to accurately capture LDES value, and inter-annual variability is likely to become more important. In line with other studies, we also find that employing the period-linking methodology to ensure energy can be transferred by LDES between representative periods is crucial to capturing the full value of LDES. For example, even when modeling a full year with 52 unlinked representative weeks, the full value of LDES cannot be captured, since such a model structure does not allow LDES to shift energy between weeks. 

Secondly, we find that the value of LDES is minimally sensitive to spatial resolution, although we expect that this finding is sensitive to the specific interconnection cost methodology used. We also wish to emphasize that other studies have found that other outcomes of interest, such as total system cost and the capacity build decisions for generators, are more sensitive to spatial resolution \cite{martinez-gordon_review_2021, krishnan_evaluating_2016, frysztacki_inverse_2023, frysztacki_strong_2021}.

One important limitation of this study is that we only examine one weather year (2012). Other studies have found that differences in capacity expansion model outcomes between weather years can be significant, particularly for very high renewables systems, such as the one we model \cite{pfenninger_dealing_2017}, and also that LDES sizing specifically is highly dependent on weather year \cite{pezza_sizing_2023}. This indicates both that our results may differ under a different weather year, and that interannual energy shifting may be a source of value that we do not capture in this study. However, we do capture the value of LDES from contributions to the capacity reserve margin, which is intended to ensure reliability across system conditions not explicitly modeled. Additionally, we expect that the general findings of this study about the convergence properties of LDES value as a function of model resolution would hold had we modeled other weather years, though the specific estimated value may differ. Subsequent work should examine the value of LDES across multiple weather years, including through comparing time sampling methods that select representative periods from across multiple years rather than a single year as in this study. 

\section{Conclusions}
\label{sec:conclusions}

Our results indicate that temporal resolution and temporal model structure are crucial considerations for energy system modelers wishing to model LDES. This is important because accurately modeling LDES is critical to accurately capturing the costs of deep decarbonization of the electric power sector, as demonstrated in Figure \ref{fig:ldes_decarb_cost}. Given our findings as well as the broader academic literature, we propose several best practices for modeling long duration storage in deeply decarbonized energy systems:

\begin{enumerate}
\item If using representative periods, ensure that the periods are linked, so that LDES resources can shift energy between periods and also discharge for their full duration. The period-linking methodology developed by Kotzur et al. \cite{kotzur_time_2018} is a current best practice for doing this, and it is implemented in the GenX open source capacity expansion model \cite{jenkins_enhanced_2017} with a slight improvement as discussed above.
\item Understand that the value of LDES (and as a result the optimal build of LDES, in cases where a cost is assigned and capcaity optimized) is highly dependent on the cost and availability of other resources available in the system. As a result, LDES value is not a fixed attribute, but rather a highly uncertain emergent property dependent on many other parameters in the system, especially the (as yet highly uncertain) costs of other ``clean firm'' technologies in the context of a low or zero carbon constraint.
\item To determine the proper amount of temporal resolution necessary to capture the full value of LDES, consider what other resources are available in the system. If LDES is serving as the sole alternative for a high fixed cost, low variable cost firm resource like advanced nuclear, then a high amount of temporal resolution is likely not necessary to capture the value of LDES (although the presence of linked periods and an accurate capacity reserve margin formulation is still necessary, so that LDES has the ability to displace such firm resources). If, on the other hand, LDES has the potential to serve a significant energy arbitrage role, then a higher amount of temporal resolution is likely to be necessary in order to accurately capture the distribution of energy price spreads, and interannual variability is likely to play a role as well. In our study, we only found convergence to within around 10\% of the true value starting at around 6000 operational hours of temporal resolution in cases where LDES derives significant value from energy arbitrage, although this specific number is likely dependent on the weather year.
\end{enumerate}

\section{Data availability}

The data behind the figures in this study are available upon request from the authors. All input data was derived from publicly available US energy system data and processed using the PowerGenome tool \cite{schivley_powergenomepowergenome_2022}.

\section{Declaration of Interests}
Jesse D. Jenkins is part owner of DeSolve, LLC, which provides techno-economic analysis and decision support for clean energy technology ventures and investors (a full list of clients can be viewed at http://linkedin.com/in/jessedjenkins). He serves on the advisory board of Eavor Technologies Inc., Rondo Energy, and Dig Energy and has an equity interest in each company. He also serves as a technical advisor to MUUS Climate Partners and Energy Impact Partners. The remaining authors declare no competing interests.

\section{Acknowledgements}

This research was supported by funding from the Princeton Zero-carbon Technology Consortium, which is supported by unrestricted gifts from GE, Google, Clear Path and Breakthrough Energy. The authors are also pleased to acknowledge that the work reported on in this paper was substantially performed using the Princeton Research Computing resources at Princeton University which is consortium of groups led by the Princeton Institute for Computational Science and Engineering (PICSciE) and Office of Information Technology's Research Computing.

\appendix
\renewcommand{\thesection}{Appendix \Alph{section}}

\section{Supplementary results}
\label{sec:extra_results}

\begin{figure}[htbp]
  \centering
  \includegraphics[width=1\textwidth]{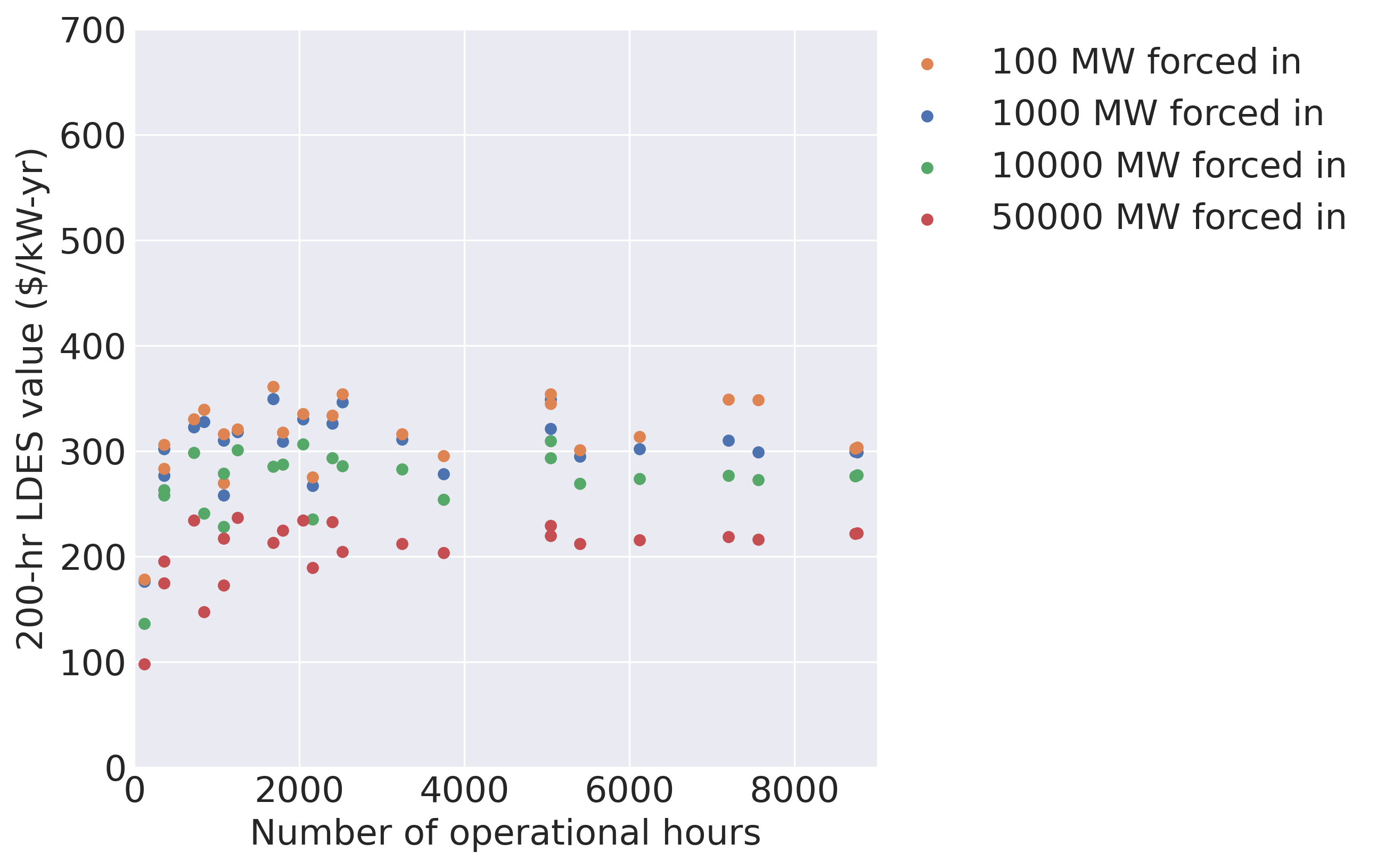}
  \caption{LDES value as a function of number of operational hours included, for varying amount of LDES forced in.}
  \label{fig:impact_of_amount_forced_in}
\end{figure}

Figure \ref{fig:impact_of_amount_forced_in} shows the impact of the capacity of LDES required on the LDES value trend. These results tell us whether the trends hold at higher penetrations of LDES. The results indicate a declining marginal value of LDES, as expected, but the variation in estimated value is largely consistent as temporal resolution increases, indicating that a `sufficient' temporal resolution is largely independent of the LDES penetration level. 

\begin{figure}[htbp]
  \centering
  \includegraphics[width=1\textwidth]{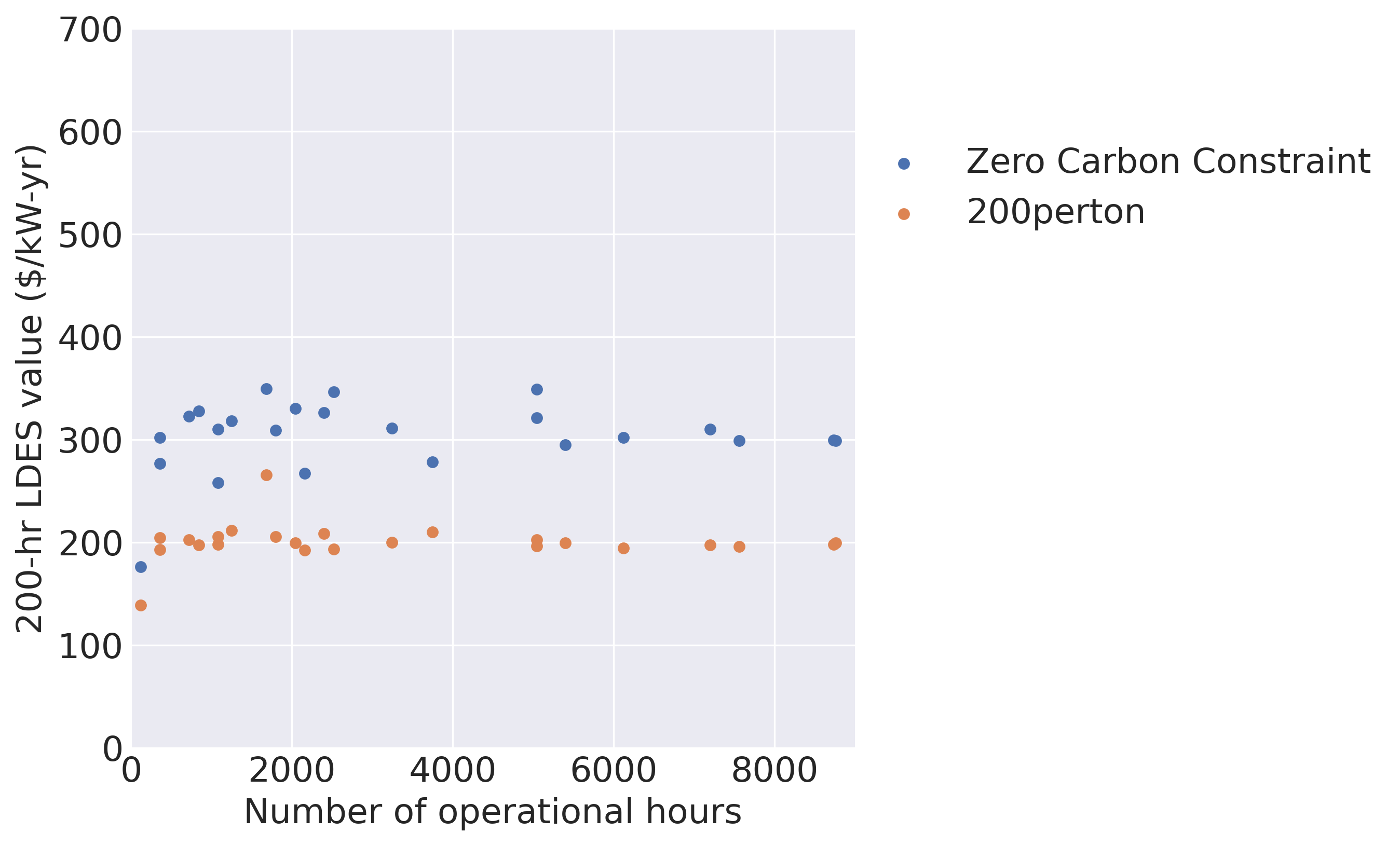}
  \caption{LDES value as a function of number of operational hours included, with both a zero carbon constraint and a \$200/ton carbon price.}
  \label{fig:emissions_impact}
\end{figure}

Figure \ref{fig:emissions_impact} shows the impact of relaxing the zero emissions constraint, and instead setting a \$200/ton carbon price. In this case, conventional gas power plants remain in the system but incur high marginal costs (\$120-180/MWh) due to the carbon emissions penalty, although this is still cheaper than the variable cost of Zero Carbon CTs in prior examples (around \$200/MWh). The value of LDES is consequently lower in cases with a carbon price, but estimated LDES value is still noisy for very low temporal resolution. These results show that the findings of this study are likely to apply not only to cases with a zero emissions constraint, but also to low- but not absolute-zero carbon cases.

\begin{figure}[htbp]
  \centering
  \includegraphics[width=1\textwidth]{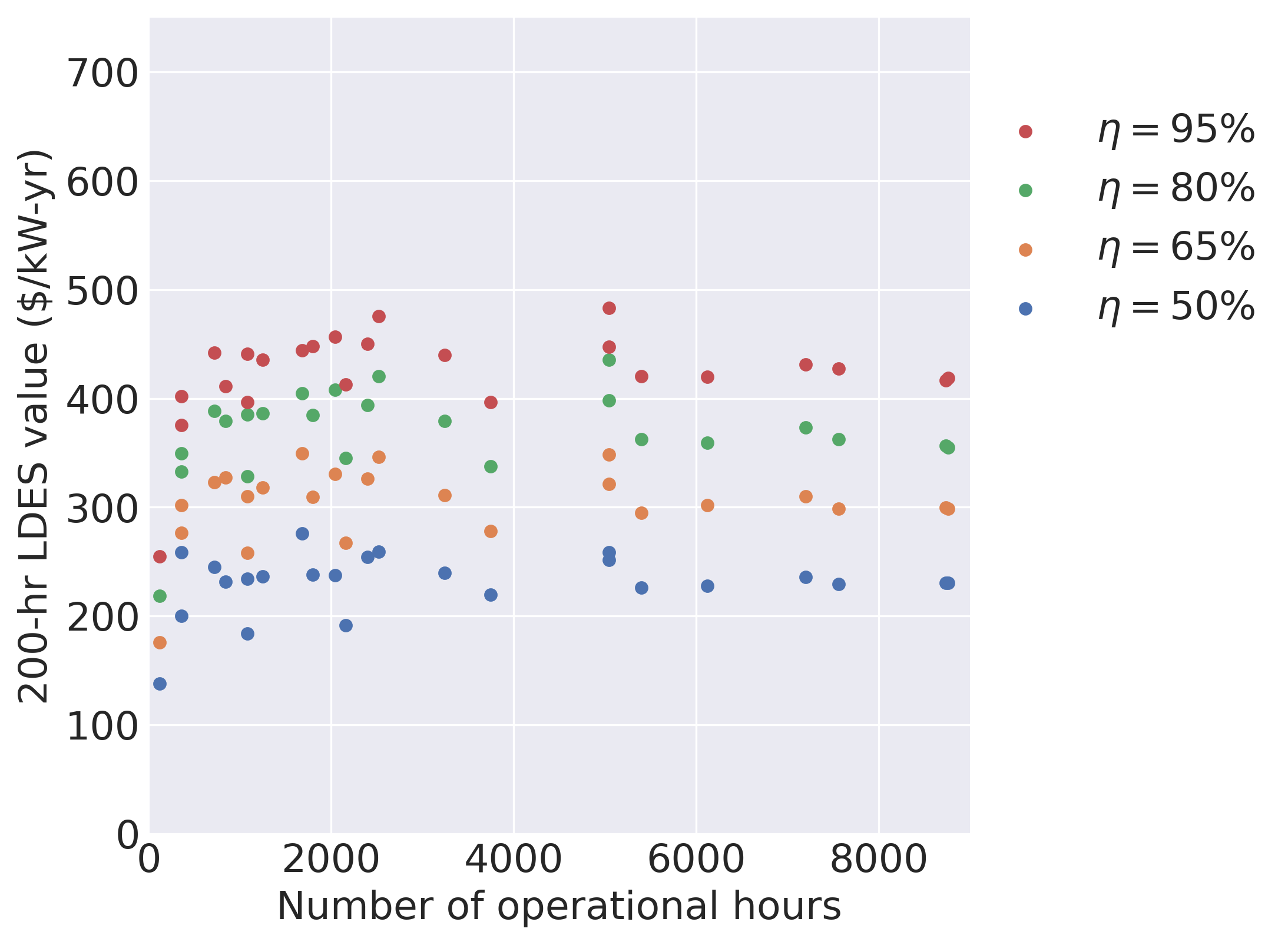}
  \caption{LDES value as a function of number of operational hours included, for varying round-trip efficiency values.}
  \label{fig:rte}
\end{figure}

Figure \ref{fig:rte} shows the impact of round-trip efficiency (RTE) on the value of LDES. The results indicate that the trend in value as a function of temporal resolution is not sensitive to round-trip efficiency; rather, the value is more or less shifted down for lower round-trip efficiency.

\begin{figure}[htbp]
  \centering
  \includegraphics[width=1\textwidth]{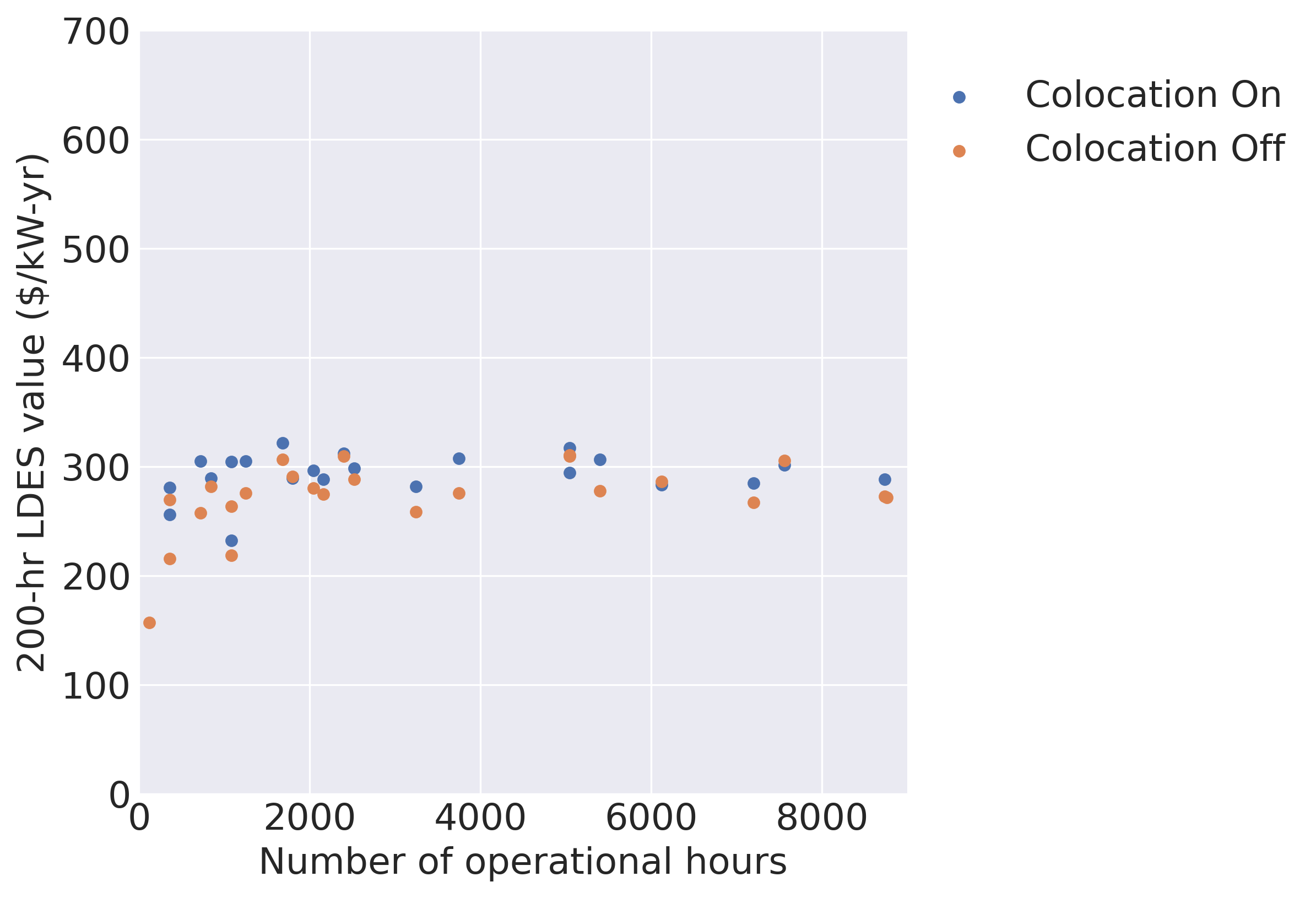}
  \caption{LDES value as a function of number of operational hours included, with and without the colocation of LDES with renewables.}
  \label{fig:colocation_impact}
\end{figure}

Figure \ref{fig:colocation_impact} shows the impact of allowing LDES to be co-located with renewables on the LDES value trend. Through co-location, LDES is able to reduce the inverter and grid connection costs for these renewable resources and may derive higher value than stand-alone storage installations. The results indicate that co-location increases the value of LDES slightly, and leads to a noisier value trend, as the value is more sensitive to noise in the time domain reduction when co-location is allowed. However the overall trends in value as a function of temporal resolution remain the same.

\begin{figure}[htbp]
  \centering
  \includegraphics[width=1\textwidth]{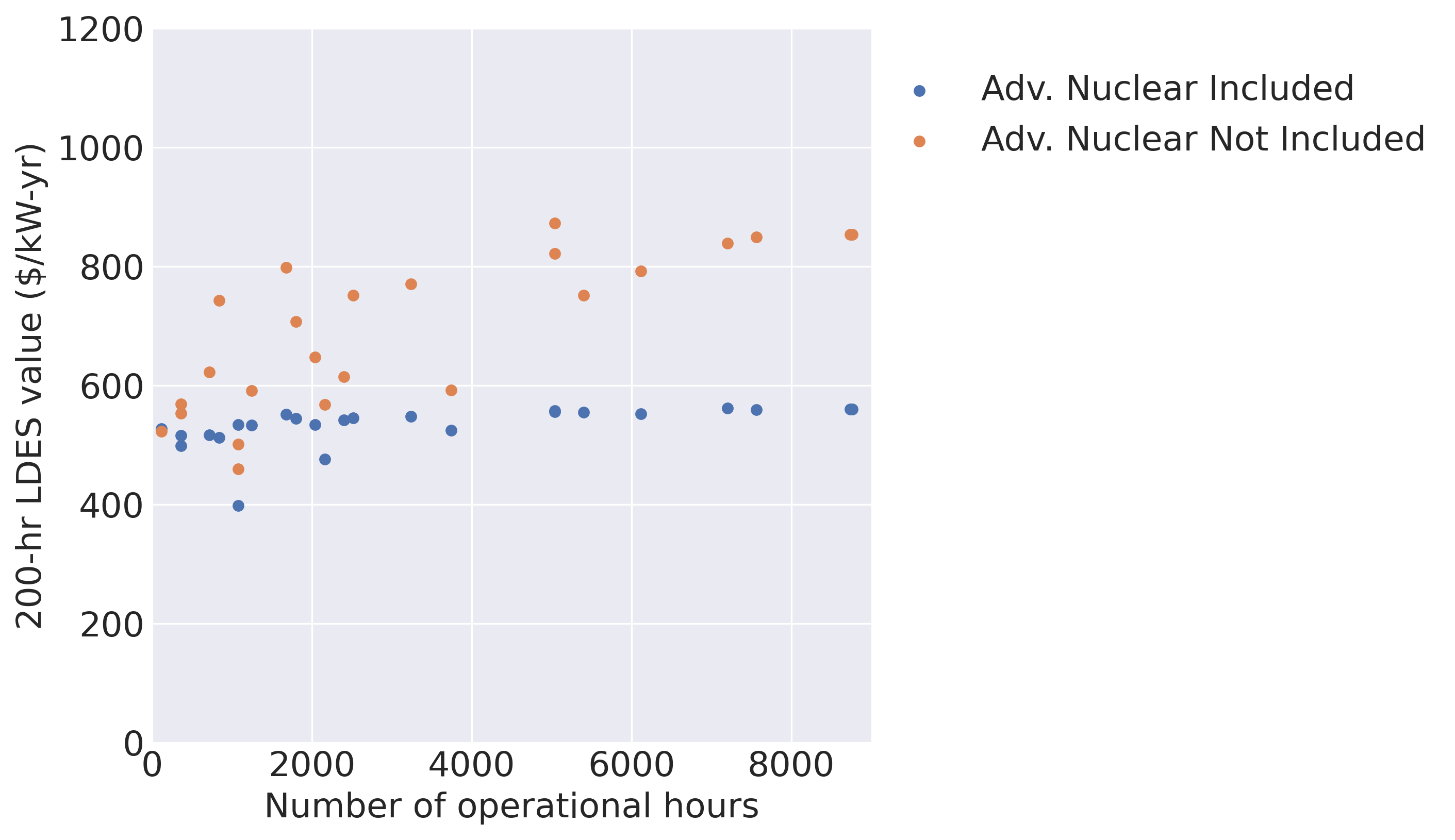}
  \caption{LDES value as a function of number of operational hours included, with and without the Advanced Nuclear resource included. None of these cases allow the Zero Carbon CT resource to be built.}
  \label{fig:impact_of_no_nuclear_12z}
\end{figure}

Figure \ref{fig:impact_of_no_nuclear_12z} shows the impact of disallowing new advanced nuclear resources (in all cases shown in this plot, the Zero Carbon CT resource is not included). These results indicate that without Advanced Nuclear (here, renewables and Li-ion batteries are the only available new resources), the value of LDES becomes significantly noisier and more dependent on the specific time domain reduction used, and does not show a clear convergence trend (the most temporally resolved cases were not computationally feasible for this sensitivity). However, it has been well documented that achieving absolute zero emissions without any clean firm resources or long duration storage is not economically feasible or realistic \cite{jenkins_getting_2018}, so we opt not to focus on this case.

\begin{figure}[htbp]
  \centering
  \includegraphics[width=1\textwidth]{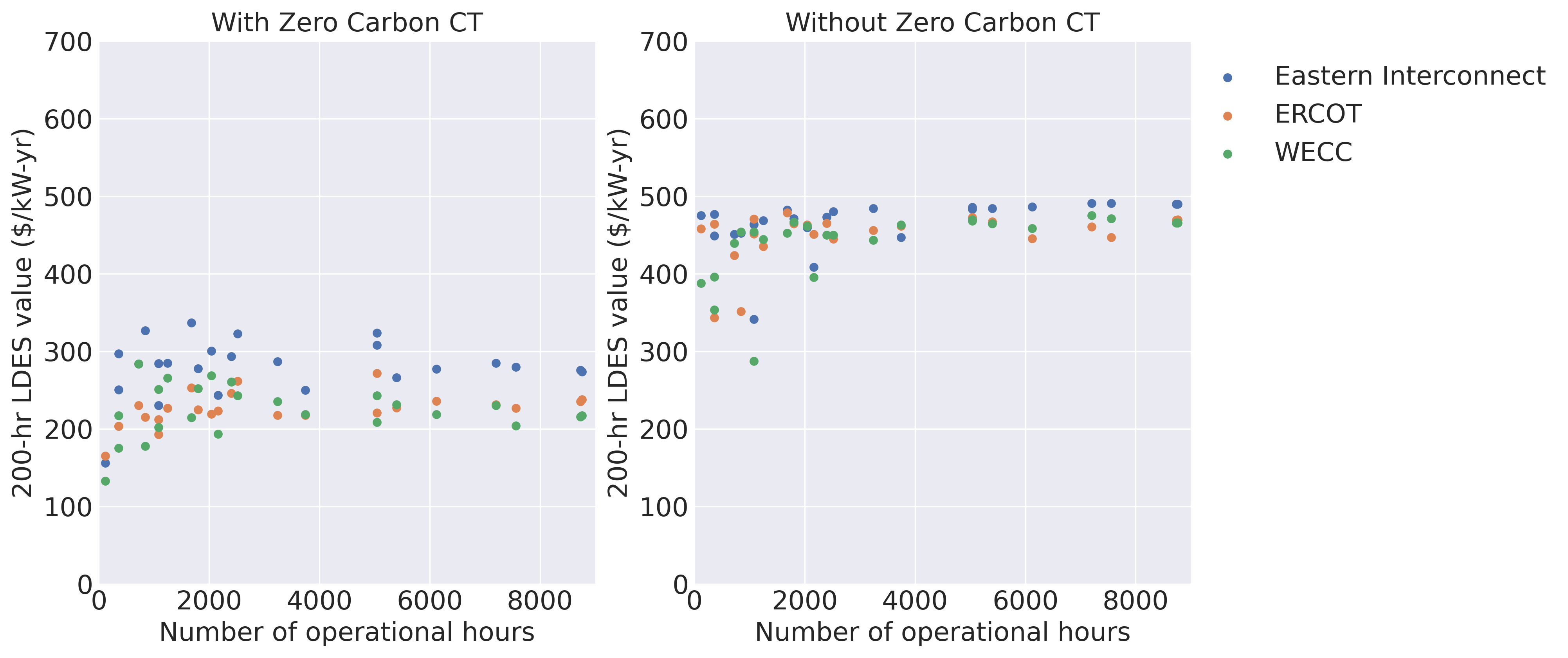}
  \caption{LDES value as a function of number of operational hours included, broken down by interconnection, with and without the Zero Carbon CT resource included.}
  \label{fig:zone_trends}
\end{figure}

Figure \ref{fig:zone_trends} shows the breakout of LDES value by interconnection, with and without the Zero Carbon CT resource included. These results show that the key study findings do not vary by interconnection, indicating that the results are likely robust to different grid mixes. When LDES value is broken out by interconnection, more noise in the value as a function of operational hours is shown, but the qualitative trends are the same.

\begin{figure}[htbp]
  \centering
  \includegraphics[width=1\textwidth]{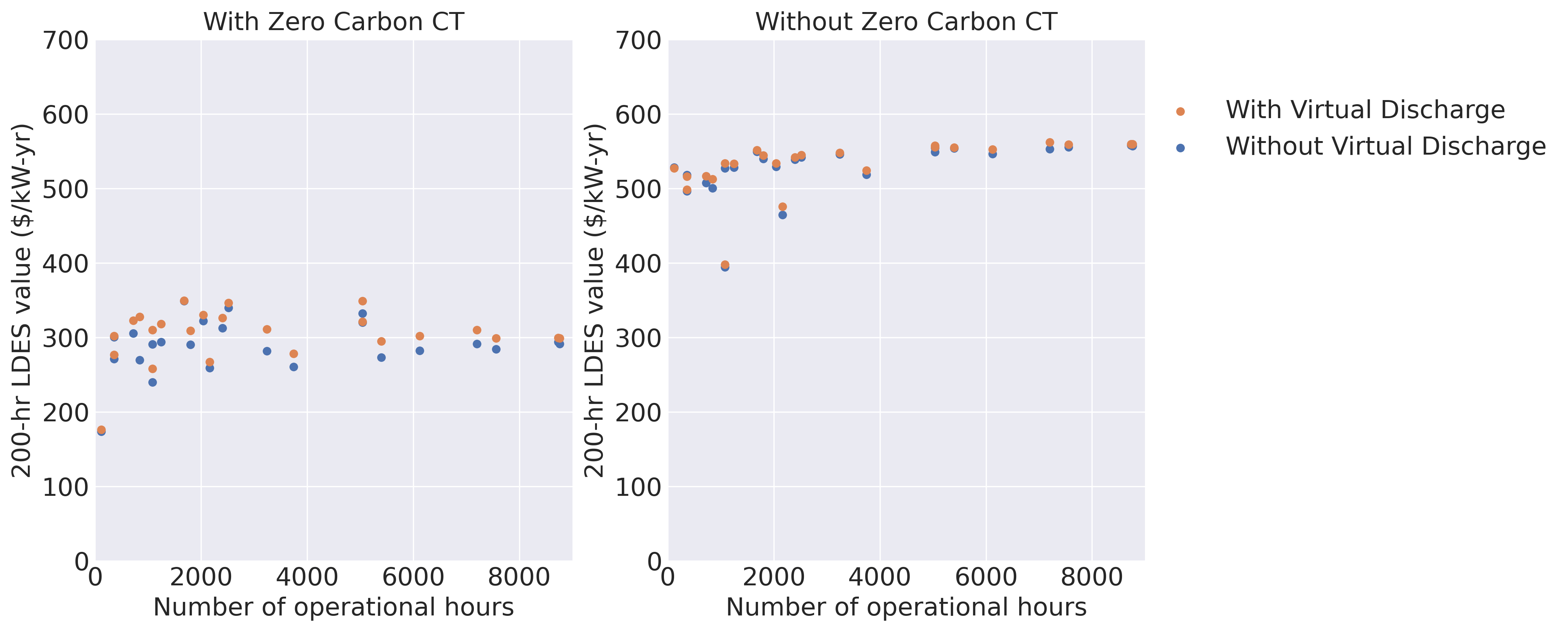}
  \caption{LDES value as a function of number of operational hours included, for the base case with Zero Carbon CT included, with and without the storage ``virtual discharge'' capacity reserve margin formulation turned on.}
  \label{fig:with_without_vdis}
\end{figure}

Figure \ref{fig:with_without_vdis} shows the impact of including the novel capacity reserve margin formulation described in this paper that allows storage to contribute to the capacity reserve margin via ``virtual discharge.'' This figure shows that the impact of including this formulation on the results is minimal; however, it can be seen that in the case with the Zero Carbon CT resource included, including this formulation leads to a value that is slightly higher. We expect that this occurs because in this case, LDES obtains its value both from energy arbitrage and from displacing capacity resources (see Figure \ref{fig:main}), so including the formulation allows for a potentially more accurate estimation of the ability of LDES to capture both of these value streams at the same time. On the other hand, when the Zero Carbon CT resource is not included, LDES obtains its value almost entirely from displacing capacity resources, so it is able to capture this value stream without needing the virtual discharge.

\section{Capacity reserve margin formulation details}
\label{sec:prm}

Expanding on the description in section \ref{sec:capres}, we now describe our novel capacity reserve margin formulation that allows storage to contribute towards the reserve margin without needing to generate. As previously described, GenX uses a capacity reserve margin formulation that endogenously calculates resource contributions toward the capacity reserve margin, and previously, storage resources were unique in their limitation of not being able to contribute towards the capacity reserve margin unless they actually generate. This is distinct from thermal and variable renewable resources which can contribute toward the margin simply by having surplus available capacity. This limitation was intentional, as storage resources need to not only have power capacity but also energy capacity in order to generate, so it is not straightforward to account for the ability of storage to provide additional energy beyond the dispatch in the model.

We solve this problem by developing a novel formulation that allows storage resources to contribute to the capacity reserve margin beyond their actual generation through a set of ``virtual'' charge, discharge, and state of charge decision variables. These variables reflect charging and discharging that could occur based on the available energy in the system. Importantly, in our formulation, virtual discharge is backed up by ``real'' state of charge in the storage resource, not just an imaginary or shadow set of operations. This acts to prevent double counting of capacity when the reserve margin is binding in more than one hour.

In our formulation, variables for virtual charge, $\Pi^{CRM}_{o,z,t}$, and virtual discharge, $\Theta^{CRM}_{o,z,t}$, are created to represent contributions that a storage device makes to the capacity reserve margin without actually generating power. The overall contribution of storage devices to the system's capacity reserve margin in a timestep $t$ is equal to $\sum_{y \in \mathcal{O}} \epsilon_{y,z,p}^{CRM} \times \left(\Theta_{y,z,t} + \Theta^{CRM}_{o,z,t} - \Pi^{CRM}_{o,z,t} - \Pi_{y,z,t} \right)$, and includes both actual and virtual charge and discharge. For storage technologies with symmetric charge and discharge capacity (all $o \in \mathcal{O}^{sym}$), charge rate, $\Pi_{o,z,t}$, and virtual charge rate, $\Pi^{CRM}_{o,z,t}$, are jointly constrained by the total installed power capacity, $\Omega_{o,z}$. Since storage resources generally represent a `cluster' of multiple similar storage devices of the same type/cost in the same zone, GenX permits storage resources to simultaneously charge and discharge (as some units could be charging while others discharge), with the simultaneous sum of charge, $\Pi_{o,z,t}$, discharge, $\Theta_{o,z,t}$, virtual charge, $\Pi^{CRM}_{o,z,t}$, and virtual discharge, $\Theta^{CRM}_{o,z,t}$, also limited by the total installed power capacity, $\Delta^{total}_{o,z}$. This constraint is as follows:
\begin{equation}
\begin{aligned}
&  \Pi_{o,z,t} + \Pi^{CRM}_{o,z,t} + \Theta_{o,z,t} + \Theta^{CRM}_{o,z,t} \leq \Delta^{total}_{o,z} & \quad \forall o \in \mathcal{O}^{sym}, z \in \mathcal{Z}, t \in \mathcal{T}
\end{aligned}
\end{equation}

In our formulation, storage resources maintain the standard state of charge tracking constraints, but are given additional constraints to track the virtual state of charge. The standard state of charge tracking is accomplished via the following two constraints, which track the state of charge of the storage resources at the end of each timepoint, relating the volume of energy stored at the end of the timepoint, $\Gamma_{o,z,t}$, to the state of charge at the end of the prior timepoint, $\Gamma_{o,z,t-1}$, the charge and discharge decisions in the current timepoint, $\Pi_{o,z,t}, \Theta_{o,z,t}$, and the self discharge rate for the storage resource (if any), $\eta_{o,z}^{loss}$.  The first of these two constraints enforces storage inventory balance for interior time steps $(t \in \mathcal{T}^{interior})$, while the second enforces storage balance constraint for the initial time step $(t \in \mathcal{T}^{start})$.
\begin{equation}
\begin{aligned}
	\Gamma_{o,z,t} &= \Gamma_{o,z,t-1} - \frac{1}{\eta_{o,z}^{discharge}}\Theta_{o,z,t} + \eta_{o,z}^{charge}\Pi_{o,z,t} - \eta_{o,z}^{loss}\Gamma_{o,z,t-1}  \\
    & \forall o \in \mathcal{O}, z \in \mathcal{Z}, t \in \mathcal{T}^{interior}\\
	\Gamma_{o,z,t} &= \Gamma_{o,z,t+\tau^{period}-1} - \frac{1}{\eta_{o,z}^{discharge}}\Theta_{o,z,t} + \eta_{o,z}^{charge}\Pi_{o,z,t} - \eta_{o,z}^{loss}\Gamma_{o,z,t+\tau^{period}-1} \\
    & \forall o \in \mathcal{O}, z \in \mathcal{Z}, t \in \mathcal{T}^{start}
\end{aligned}
\end{equation}
In our virtual discharge formulation, the following two constraints are added, which track the relationship between the virtual charge, $\Pi^{CRM}_{o,z,t}$, and virtual discharge, $\Theta^{CRM}_{o,z,t}$, variables and a third variable, $\Gamma^{CRM}_{o,z,t}$, representing the amount of state of charge that must be held in reserve to enable these virtual capacity reserve margin contributions, ensuring that the storage device could deliver its pledged capacity if called upon to do so without affecting its operations in other timepoints. $\Gamma^{CRM}_{o,z,t}$ is tracked similarly to the devices overall state of charge based on its value in the previous timepoint and the virtual charge and discharge in the current timepoint. Unlike the regular state of charge, virtual discharge $\Theta^{CRM}_{o,z,t}$ increases $\Gamma^{CRM}_{o,z,t}$ (as more charge must be held in reserve to support more virtual discharge), and $\Pi^{CRM}_{o,z,t}$ reduces $\Gamma^{CRM}_{o,z,t}$.
\begin{equation}
\begin{aligned}
	  \Gamma^{CRM}_{o,z,t} &= \Gamma^{CRM}_{o,z,t-1} + \frac{1}{\eta_{o,z}^{discharge}}\Theta^{CRM}_{o,z,t} - \eta_{o,z}^{charge}\Pi^{CRM}_{o,z,t} - \eta_{o,z}^{loss}\Gamma^{CRM}_{o,z,t-1}  \\
    & \forall o \in \mathcal{O}, z \in \mathcal{Z}, t \in \mathcal{T}^{interior}\\
	  \Gamma^{CRM}_{o,z,t} &= \Gamma^{CRM}_{o,z,t+\tau^{period}-1} + \frac{1}{\eta_{o,z}^{discharge}}\Theta^{CRM}_{o,z,t} - \eta_{o,z}^{charge}\Pi^{CRM}_{o,z,t} - \eta_{o,z}^{loss}\Gamma^{CRM}_{o,z,t+\tau^{period}-1}  \\
    & \forall o \in \mathcal{O}, z \in \mathcal{Z}, t \in \mathcal{T}^{start}
\end{aligned}
\end{equation}
Another way to think of this formulation is that the virtual state of charge represents a portion of the real state of charge that has been used by the virtual discharge. Thus, in order to virtual discharge, a storage resource must increase its state of charge with real charging (supported by other resources in the system). It is only when the virtual discharge happens that the virtual state of charge comes into play. When virtual discharge happens, the virtual state of charge increases-- reflecting the portion of the real state of charge that has been used up. In order to unlock this portion of the real state of charge for real discharge, storage must perform virtual charging. This dynamic is crucial to understanding why the charging and discharging signs are are flipped for virtual state of charge in our formulation.

In order to ensure that the virtual state of charge is a subset of the real state of charge, we constrain the real state of charge $\Gamma_{o,z,t}$ to be greater than or equal to the virtual state of charge $\Gamma^{CRM}_{o,z,t}$:

\begin{equation}
\begin{aligned}
	&  \Gamma_{o,z,t} \geq \Gamma^{CRM}_{o,z,t} 
\end{aligned}
\end{equation}

Finally, the state of charge $\Gamma_{o,z,t}$ is constrained to be less than the installed energy storage capacity, $\Delta^{total, energy}_{o,z}$, and the maximum combined discharge and virtual discharge rate for storage resources, $\Pi_{o,z,t} + \Pi^{CRM}_{o,z,t}$, is constrained to be less than the state of charge at the end of the last period, $\Gamma_{o,z,t-1}$:
\begin{equation}
\begin{aligned}
	&  \Gamma_{o,z,t} \leq \Delta^{total, energy}_{o,z} & \quad \forall o \in \mathcal{O}, z \in \mathcal{Z}, t \in \mathcal{T}\\
	&  \Theta_{o,z,t} + \Theta^{CRM}_{o,z,t} \leq \Gamma_{o,z,t-1} & \quad \forall o \in \mathcal{O}, z \in \mathcal{Z}, t \in \mathcal{T}
\end{aligned}
\end{equation}

\begin{figure}[htbp]
  \centering
  \includegraphics[width=1\textwidth]{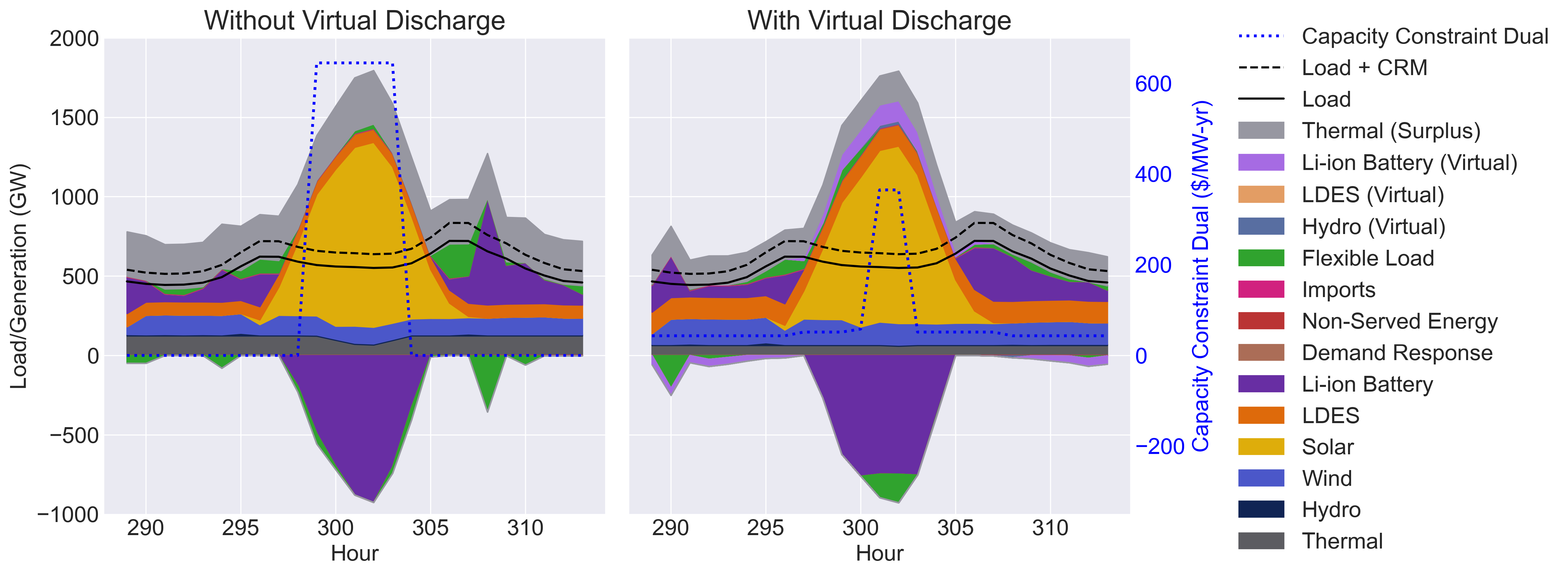}
  \caption{Dispatch on Eastern Interconnect peak day in a zero carbon case, with and without virtual discharge turned on.}
  \label{fig:vdis}
\end{figure}

The behavior of this formulation is illustrated in Figure \ref{fig:vdis}, which shows the dispatch for the Eastern Interconnect in a zero carbon case during the day when the shadow price of the capacity reserve margin constraint is highest, with and without the virtual discharge formulation enabled. This peak day is not necessarily the day with the highest system load, but rather the ``net'' peak that is most challenging to meet with zero carbon resources. This case is taken from the cases used in the study, except LDES is given an arbitrary cost of \$200/kW-yr (for illustrative purposes) and not forced in, so that the model is able to build as much LDES as it needs.

The figure shows that, in the case without the virtual discharge enabled, the model builds a large amount of surplus thermal capacity, and does not dispatch it, during this peak day so that the capacity reserve margin constraint can be met. When the virtual discharge formulation is turned on, the model is able to reduce this surplus thermal capacity significantly, and instead rely on virtual discharge to partially meet the capacity reserve margin requirements. This causes the shadow price of the capacity reserve margin constraint to decrease, as can be seen from the plot, and also reduces overall system costs by about 1\%. This case illustrates that the virtual discharge formulation enables storage resources to be treated on a level playing field with thermal resources by allowing them to contribute to the capacity reserve margin without actually dispatching, reducing system costs in the process.

It is important to note that, while this formulation represents an improvement from the previous GenX formulation which did not allow storage to contribute to the capacity reserve margin without dispatching, the GenX capacity reserve margin formulation still involves an approximation of capacity value that is based only on the weather conditions modeled. It is likely that basing the capacity contribution of resources on a methodology that evaluates performance across many possible weather conditions is a more robust way to ensure system reliability, but this topic is an important area for future research.




\bibliography{bibliography.bib}

\end{document}